\documentclass{elsart}

\usepackage{graphicx,amsfonts,amsmath,natbib}
\usepackage{natbib}
\usepackage{natbib,color}
\usepackage{amsmath}
\usepackage{amsfonts}
\usepackage{epsfig}
\usepackage{makeidx}
\usepackage{rotating}
\usepackage{subfigure}
\usepackage{theorem}

\def\beq{\begin{equation}}
\def\eeq{\end{equation}}

\def\p{\partial}
\def\({\left(}
\def\){\right)}

\def\[{\left[}
\def\]{\right]}

\newcommand{\Rey}{{Re}}
\newcommand{\D}{{\mathcal{D}}}

\newcommand{\lb}[1]{{\color{black} #1}}

\usepackage{amssymb}

\begin{document}
\sloppy

\begin{frontmatter}


\title{The lift-up effect:\\ the linear mechanism behind transition and turbulence in shear flows}
\author{Luca Brandt}
\ead{luca@mech.kth.se}
\address{Linn{\'e} Flow Centre and SeRC (Swedish e-Science Research Centre),\\ KTH Mechanics, SE-100 44, Stockholm, Sweden }

\begin{abstract}

The formation and amplification of streamwise velocity perturbations induced by cross-stream disturbances is ubiquitous in shear flows. This disturbance growth mechanism, so neatly identified by Ellingsen and Palm in 1975, is 
a key process in transition to turbulence and self-sustained turbulence.
In this review, we first present the original derivation and early studies and then discuss the non-modal growth of streaks, the result of the lift-up process, in transitional and turbulent shear flows.
In the second part, the effects on the lift-up process of additives in the fluid and of a second phase are discussed and new results presented with emphasis on particle-laden shear flows. For all cases considered, we see the lift-up process to be a very robust process, always present as a first step in subcritical transition.

\end{abstract}




\end{frontmatter}


\section{Introduction}

\subsection{"Stability of linear flow"}

This is the title of a research note in Physics of Fluids of less than two pages published in 1975 by Ellingsen and Palm. \nocite{ellingsen-Palm} In this work, the authors identify a linear mechanism responsible for the amplification of fluctuations in shear flows. 
In their own words, {\it a finite disturbance independent of the streamwise coordinate may lead to instability of linear flow, even though the basic velocity does not possess  any inflection point.}
This mechanism, later denoted lift-up effect, is a key process in the laminar-turbulent transition in shear flows and in fully developed turbulence, as will be discussed in this review.

At the time of their note,  
the main general results for the linear stability of shear flows were Rayleigh, Fj\o rtoft and Howard criteria \citep{Drazin81}. Rayleigh's criterion states that a necessary condition for the instability of a parallel shear flow is that the basic velocity profile has an inflection point \citep{Ray}.  Later \cite{Fjo50} showed that the vorticity needs to have a maximum at the inflection point. \cite{Howard} proved that the complex phase velocity of an exponential wave must lie within a semi-circle having a diameter equal to the difference between the largest and the smallest velocity of the parallel base flow.
These theorems are valid in an inviscid and not stratified fluid and were obtained by considering two-dimensional infinitesimal perturbations, i.e. directly from the linearized Rayleigh equation for the stability of a parallel shear flow. Squire's theorem (1933) 
\nocite{Squire1933}
states that two-dimensional disturbances are the first to become unstable in parallel shear flows and thus they determine the critical Reynolds number; this had restricted the stability analyses to two-dimensional normal modes (exponential growth or decay of periodic waves).

Ellingsen and Palm's fundamental contribution is to show that three-dimensional disturbances may lead to an instability other than modal, independent of the existence of an inflection point. They note how this instability can be responsible for transition to turbulence and acknowledge previous suggestions by H\o iland (referring to some unspecified lecture notes), who, however, {\it did not draw full conclusions from his idea.}  Indeed, this new  mechanism is able to explain transition in subcritical conditions or in stable flows as in the case of pipe flow \citep{Hof04}.

We will shortly outline here the original derivations and denote a parallel velocity profile as $\mathbf{U}=(U,V,W)=(U(y),0,0)$ where $U$ is the streamwise velocity component and $y$ and $z$ the cross-stream coordinates.  Considering an inviscid, incompressible and not stratified flow bounded by two parallel planes and a disturbance independent of the streamwise coordinate $x$, the equation for the streamwise component of the momentum and for the streamwise vorticity component reduce to
\begin{equation}\label{eq:first}
\frac{D u}{Dt}=0; \quad \frac{D \xi}{Dt}=0.
\end{equation}
Introducing a streamfunction $\Psi$ for the cross-stream components,
\begin{equation}\nonumber
v=\frac{\partial \Psi}{\partial z}, w=-\frac{\partial \Psi}{\partial y};
\end{equation}
and linearizing one obtains 
\begin{equation}\label{eq:2}
\frac{\partial u}{\partial t} + v\frac{d U}{d y}=0
\end{equation}
for the streamwise disturbance velocity and 
\begin{equation}\label{eq:1}
\frac{\partial }{\partial t}  \nabla_1^2 \Psi =0
\end{equation}
for the cross-stream flow, where $\nabla_1^2$ is the two-dimensional Laplacian.
From eq.(\ref{eq:1}), we see that the cross-stream velocity components are independent of time, i.e.\ a streamwise independent perturbation  $v$ will not grow or decay in an inviscid flow. Equation (\ref{eq:2}) can be integrated
\begin{equation}\label{eq:3}
u=u(0)-v \frac{d U}{d y} t
\end{equation}
to show that the perturbation $u$ grows linearly in time, from which also the name of algebraic inviscid instability. It is hence shown that any shear flow $U(y)$ is unstable to streamwise independent disturbances in the cross-stream velocity components.

This first part of the original paper, based on a linear analysis, is probably the most known and commonly used as a reference for the optimal transient growth of $x$-independent perturbations in viscous flows, see \cite{Schmid2001} and discussion below.  Indeed, we will see that infinitely long streamwise vortices are the most dangerous initial conditions in shear flows: they lead to the formation of streamwise streaks, elongated regions of positive and negative streamwise velocity, by redistributing streamwise momentum across the shear layer.

Ellingsen and Palm, in addition, show that the equations (\ref{eq:first}) can be solved also for finite-amplitude perturbations. The conservation of streamwise vorticity  can be re-written as
\begin{equation}\label{eq:nonlin}
\frac{\partial }{\partial t}  \nabla_1^2 \Psi  + \frac{\partial \Psi}{\partial z} \frac{\partial}{\partial y} \nabla_1^2 \Psi - \frac{\partial \Psi}{\partial y} \frac{\partial}{\partial z} \nabla_1^2 \Psi =0,
\end{equation}
which admits solution of the form 
\begin{equation}\label{eq:6}
  \nabla_1^2 \Psi = f(\Psi), \quad \frac{\partial \Psi}{\partial t} =0,
\end{equation}
with $f$ an arbitrary function. If $f$ is  a linear function, the cross-stream motion is represented by a set of closed streamlines.
The conservation of momentum in the streamwise direction then implies that the velocity $u$ is conserved during the motion along these closed streamlines. {\it A fluid particle in its orbit in the x-y plane will, therefore, have a $u$ velocity equal to the value of the basic flow at the initial position of that particle.} This value will be different from the initial local value of $u$, the more the larger the vertical particle displacement in a homogeneous shear. As the period can be different along different streamlines, the motion is aperiodic with a complete redistribution of streamwise momentum. This is independent of the initial disturbance amplitude  and may lead to large velocity gradients that, in turn, can support new instabilities: {\it It is possible of course that the developed motion is unstable. Owing to the large vorticity concentrations this indeed seems very likely so that the motion already discussed is valid only for a short span of time}.  This is indeed what happens in the case of secondary streak instability, where an inflectional type of instability develops on the regions of largest vorticity induced by streamwise elongated perturbations \citep{sreddy:1998,paul-luca,lucaparis}. However, the vertical displacement of fluid particles by the cross-stream momentum is not observed  only for a short time, as cautiously stated by Ellingsen and Palm. This is a key ingredient not only for the breakdown to turbulence but also in the dynamics of wall-bounded turbulence, as we will show in this review.

Ellingsen and Palm conclude that, despite their analysis is limited to the case of streamwise independent disturbances, the equations are valid also when the base flow has an angle with the $x$-direction, $\mathbf{U}=(U(y),0,W(y))$. For small angles, the physical mechanisms at play (cross-stream displacement of fluid particles that retain their horizontal momentum) is the same. For larger angles, however, the variations of the streamwise velocity $u$ are much smaller as the disturbance field has a component in the $x$-direction. This seminal paper ends by stating that {\it by same reasoning we obtain the result that an inviscid channel flow is always unstable for perturbations independent of the streamwise coordinate.} This explains the first stage of the \lb {subcritical} transition to turbulence in pipe flow, a problem that puzzled scientists for over a century \citep{Eckhardt2007,Mullin2011}. \lb{Indeed the lift-up effect becomes the main responsible for disturbance energy growth when no other modal instabilities are present.}

\subsection{Early inviscid studies}

In a review paper from 1969 about shear-flow turbulence, Phillips reports a previous analysis by Moffatt aiming to explore whether a disturbance can maintain itself by interactions with the mean shear \citep{moffat67,phillips69}.
Considering the interactions between middle-size eddies and a uniform shear flow, $U= S y$, Moffatt determined solutions of the linearized Navier--Stokes equations for three-periodic velocity perturbations and pressure 
\begin{equation}\label{eq:moffat1}
u_i=A_i(t) \exp [i (\mathbf{k}(t) \cdot \mathbf{x}) ];  \quad p=\pi(t) \exp [i (\mathbf{k}(t) \cdot \mathbf{x}) ] 
\end{equation}
with wavenumber
\begin{equation}\label{eq:moffat2}
\mathbf{k}(t)=(k_x, k_y, k_z)=[k_x(0), k_y(0) - St k_x(0), k_z(0)].
\end{equation}
The latter expression indicates that each Fourier component is tilted by the shear, where the lines of constant phase move closer together and rotate until they become asymptotically parallel to the planes defined by a constant value of the coordinate $y$. Moffatt also derives a dynamical equation of the velocity amplitudes $A_i$ and shows that for streamwise independent modes ($k_x=0$) the solution can be written as
\begin{equation}\label{eq:moffate}
A_x(t)=A_x(0)- St A_y(0); \, A_y(t)=A_y(0): \, A_z(t)=A_z(0).
\end{equation}
The streamwise velocity perturbation grows linearly in time if the initial disturbance has a non-zero component in the wall-normal direction, as shown by Ellingsen and Palm for a bounded shear flow and any general disturbance shape in the linear and nonlinear regime.
A superposition of periodic disturbances evolves towards a series of horizontal structures with a vanishing cross-stream velocity components and vanishingly small scales in the $y$-direction, something which would accelerate viscous dissipation.
Moffatt calculated the Reynolds stress associated to these structures and showed that the flow will asymptotically tend to one dominated by large-scale structures, independent of the $x$-coordinate. Phillipps notes in his review that the disturbance amplification computed by Moffatt corresponds to cross-stream displacement of fluid particles retaining their original streamwise momentum but this cannot explain how turbulence is sustained although there is abundant evidence of the presence of such elongated structures in wall turbulence.
Studies of homogenous-shear turbulence shed anyway light on the energy transfer among Fourier modes represented by the tilting of the disturbance and the lifting of the elongated streaks observed in turbulence.

\lb{As mentioned above, the historical basis for the paper by Ellingsen and Palm was the work on hydrodynamic stability by Palm's mentor Einar H\o iland. Remarkably, Palm's paper was his last contribution to the stability analysis of homogeneous fluids. Nobody in Norway followed up this research. This was however continued in Sweden, due to the influence of Palm's friend, M\aa rten Landahl.}
Few years later, \cite{Landahl75} studied the dynamics of shear flow turbulence and the burst events, always associated to a low-speed streak lifting from the surface and forming locally a highly inflectional velocity profile. As these are deterministic and repeatable events he carried out a mechanistic analysis based on the linearized equations and a two-scale model (triple decomposition). Landahl shows that the large-scale streamwise-velocity fluctuations produced by a localized burst elongate in the streamwise direction: his analysis shows that {\it the disturbance created by the burst will leave a "permanent scar" in the flow, convected downstream with the local flow velocity. In reality, viscosity will of course make this disturbance decay, but on a time scale much greater than the decay time of the transient wall-normal velocity disturbances produced during the burst.} Landahl predicts that the perturbations produced by the longer waves have larger speeds as they have their maxima further away from the wall, hence the streak will appear to move towards the wall to a fixed observer \citep[see the experiments in][]{Lundell03}.

\cite{Landahl75} writes the counterpart of equation(\ref{eq:2}) for the large scale turbulent motions $\tilde u'$ neglecting stress and pressure gradient
\begin{equation}
\frac{\partial \tilde u'}{\partial t} + U \frac{\partial \tilde u'}{\partial x}+ \tilde v' \frac{d U}{d y}=0,
\end{equation}
and shows that its solution is the linear analogous of Prandtl's mixing length hypothesis that each fluid particle would retain its horizontal momentum as it is displaced normal to the wall. Concluding, Landahl notes that the interpretation proposed is based on a reverse cascade of energy, from small-scale bursts to large scale turbulent fluctuations, i.e.\ the opposite of classical turbulence theory with a energy cascade to smaller and smaller scales until viscous dissipation provides a cut-off mechanism. This is indeed the peculiarity of wall-bounded turbulence that makes modeling of these flow a particularly  hard problem.

\cite{Landahl:algebraic} considers the evolution in space and time of an arbitrary three-dimensional initial disturbance and shows that a wide class of three-dimensional disturbances gives rise to a perturbation kinetic energy growing at least as fast as  linearly with time in any inviscid shear flow. This is because the size of the perturbed region grows linearly in time while the streamwise velocity disturbance does not decrease as $t \to \infty$. To show this, Landahl introduces  an average in the streamwise direction
\begin{equation}
 \mathbf{\bar u} = \int_{-\infty}^\infty  \mathbf{u} \, dx. 
\end{equation}
and shows that $\bar p=0$ and that $\bar v=\bar v_0$ is independent of time. Integration of the streamwise-averaged streamwise momentum equation (cf. eq. \ref{eq:2}) gives that
\begin{equation}
 {\bar u} = {\bar u_0} -t {\bar v_0} \frac{dU}{dy}.
\end{equation}
The fact that the integrated streamwise momentum increases linearly in time does not imply that $u$ increases, since the disturbance may and indeed does spread in time. Mathematically we retrieve the same behavior as $x$--independent disturbances \citep{ellingsen-Palm} but for a localized disturbance.
Using Schwartz's inequality and introducing the positive quantities $\gamma$ and $[U]=U_{max}-U_{min}$ where the latter two are the maximum and minimum velocity of the base flow, Landahl shows that the total integrated kinetic energy of the disturbance
\begin{equation}
E=\frac{1}{2} \int_{-\infty}^\infty  ({u}^2+{v}^2+{w}^2) \, dx > \frac{1}{2 ([U]+ 2 \Gamma)}  {\bar v_0}^2 (\frac{dU}{dy})^2 t .
\end{equation}
In other words, the total kinetic energy of a localized disturbance with $\bar v_0 \ne 0$ will grow at least linearly with time.
This results holds for asymptotically stable and unstable shear flows: any inviscid shear flow will experience the growth of three-dimensional disturbances provided $\bar v_0 \ne 0$. For flows without an inflection point this growth is associated to the streamwise velocity disturbance 
$u$ and explains the tendency of transition and turbulent shear flows to develop longitudinal streaky structures.

The asymptotic analysis in the appendix of \cite{Landahl:algebraic} shows that the velocity perturbation $u$ remains bounded as $t \to \infty$ and the streamwise extension of the disturbed region grows linearly in time: elongated streaks will therefore form in a shear flow.
The evolution of the maximum streamwise velocity was studied more recently by means of numerical simulations at increasing Reynolds number  by \cite{Lundbladh:PhD}. It is observed that the streamwise velocity amplitude growth is logarithmic in time after an initial transient whereas the energy grows linearly in time in the inviscid limit as predicted by Landahl.

\section{Linear stability and transition to turbulence}

\subsection{The effect of viscosity}

The viscous counterpart of the solution obtained in \cite{ellingsen-Palm} is derived in \cite{Gustavsson1981}.
These authors 
consider a boundary layer and perturbations with
zero streamwise dependence and write the
equation for the streamwise disturbance velocity 
\begin{equation} \label{eq:hakan}
\frac{\p u}{\p t} - \frac{1}{\Rey} \left( \frac{\p^2}{\p y^2} + 
\frac{\p^2}{\p z^2} \right) u = - U^\prime v. 
\end{equation} 
For small times, $t/\Rey \ll 1,$ 
the vertical velocity $v$ remains constant
\begin{equation} 
v(y,z,t) = v_0(y,z) + {\mathcal{O}}(t/\Rey), 
\end{equation} 
as demonstrated in \cite{Gustavsson1981} using a Fourier-Laplace  transform.
The streamwise velocity component can be obtained as the 
solution to a diffusion equation with forcing proportional to the wall-normal velocity, see equation(\ref{eq:hakan}). For small times $t/\Rey \ll
1,$ application of standard asymptotic
techniques~\cite[see][]{Gustavsson1981} gives
\begin{align} 
u(y,z,t) \sim u_0(y,z) &+ {\mathcal{O}}(t/\Rey) \nonumber \\
&- \int_0^t \left[ v U^\prime + \frac{\tau}{\Rey} \left( 
\frac{\p^2}{\p y^2} + \frac{\p^2}{\p z^2} \right) v U^\prime 
\right] \ d\tau . 
\end{align}
The first term in the integral, the only non-vanishing term for large Reynolds numbers, 
leads to an algebraic growth in time
\begin{equation} 
u(y,z,t) \sim -v_0(y,z) U^\prime t. 
\end{equation} 
This recovers the inviscid result of~\cite{ellingsen-Palm} for bounded flows.

\subsection{The disturbance transient growth} 

The viscous linear stability of parallel shear flows is described by the Orr-Somemrfeld and Squire system for the perturbation wall-normal velocity and vorticity $\eta$ \citep{Schmid2001}. Introducing wavelike
solutions of the form 
\begin{align}
v(x,y,z,t) &= \hat v(y,t)\ e^{i(\alpha x + \beta z)} \\
\eta(x,y,z,t) &= \hat \eta(y,t)\ e^{i(\alpha x + \beta z)}.
\end{align}
where $\alpha$ and $\beta$ are the streamwise and spanwise wavenumbers, the equations  governing the time evolution of any initial disturbance read 
\begin{align}
  \left[ (\frac{\partial }{ \partial t} + i\alpha U) (\D^2-k^2)
  -i\alpha U^{\prime \prime} - \frac{1}{\Rey}(\D^2-k^2)^2\right] \hat{v} &
  =  0
  \label{eq:C5:v-hat} \\ 
\left[ (\frac{\partial }{ \partial t} + i\alpha U) -
  \frac{1}{\Rey}(\D^2-k^2)\right] \hat{\eta} & =  -i\beta
  U^{\prime}\hat{v}
  \label{eq:C5:eta-hat}
\end{align}
with boundary conditions $\hat v = \D \hat v = 
\hat \eta = 0$ at solid walls and in the free stream.
In the expressions above, $k^2=\alpha^2+\beta^2$ whereas ${\mathcal D}$ and $'$ denote $\partial/\partial y$.
Introducing the vector 
\begin{equation}
\begin{pmatrix} \hat{v} \\ \hat{\eta} \end{pmatrix}
\end{equation}
the Orr-Sommerfeld and Squire equations can be written in matrix form as
\begin{equation}
  \frac{\partial }{ \partial t}  \begin{pmatrix} k^2-\D^2 & 0 \\ 
   0 & 1 \end{pmatrix} 
   \begin{pmatrix} \hat{v} \\
   \hat{\eta} \end{pmatrix} + \begin{pmatrix} {\mathcal{L}}_{OS} & 0 \\ 
   i\beta U' & {\mathcal{L}}_{SQ} \end{pmatrix} \begin{pmatrix} \hat{v} \\ 
  \hat{\eta} \end{pmatrix} = 0
\label{eq:C4:OrrSquireMatrix}
\end{equation}
where
\begin{align}
{\mathcal{L}}_{OS} &= i\alpha  U (k^2-\D^2) +
  i\alpha  U'' + \frac{1}{\Rey}(k^2-\D^2)^2 \\
{\mathcal{L}}_{SQ} &= i\alpha  U + \frac{1}{\Rey}(k^2-\D^2).
\end{align}
The solution to this system, including boundary conditions, determines the
behavior of disturbances in parallel shear flows. 
The evolution of $\hat v$ is described by the homogeneous equation
(\ref{eq:C5:v-hat}) with homogeneous boundary conditions and can be determined
once the initial data are given. In contrast,
the off-diagonal coupling term, $i\beta
U'$, in the matrix implies that the Squire equation is driven by
solutions to the Orr-Sommerfeld equation, unless $\hat v$ or $\beta$
is zero. 
Because (\ref{eq:C5:eta-hat}) is the linearized
form of the evolution equation for normal vorticity, and the forcing
term stems from the linearized vortex tilting term, the forcing mechanism is also denoted as vortex tilting.  This forcing
originates the algebraic instability uncovered by Ellingsen and Palm.

 To show the potential for transient growth of the disturbance in the viscous case, we derive the solution of the forced Squire equation \cite[see][for more details]{Schmid2001}.
This non-modal growth mechanisms may dominate over the asymptotic behavior predicted by
the eigenmodes for short times. 
For simplicity we assume that the wall-normal velocity perturbation can be described by only 
one mode of the Orr-Sommerfeld operator $\mathcal{L}_{OS}$
\begin{equation}
  \hat{v}=\tilde{v}_l\ e^{-i\alpha c_l t},
  \label{eq:C5:v-exp1}
\end{equation}
with $c_l$ the phase speed of the eigenmode.
The solution the equation for the normal vorticity
(\ref{eq:C5:eta-hat}) using (\ref{eq:C5:v-exp1})  as a forcing term consists of a homogeneous and a particular
solution
\begin{equation}
  \hat{\eta} = \hat{\eta}_{hom} + \tilde{\eta}_l^p\ e^{-i\alpha c_l t}
  \label{eq:C5:eta-hom-part}
\end{equation}
where the time dependence of the particular solution is given by the eigenvalue of $\mathcal{L}_{OS}$. 
We express the solution for $\hat{\eta}_{hom}$ and $\tilde{\eta}_l^p$
in the eigenmodes of the homogeneous part
of the normal vorticity equation, $\mathcal{L}_{SQ}$. For the homogeneous
part we have the expansion
\begin{equation}
  \hat{\eta}_{hom} = \sum_j C_j \tilde{\eta}_j\ e^{-i\alpha
  \sigma_j t} 
\label{eq:C5:eta-hom}
\end{equation}
where $\tilde{\eta}_j$ are the
Squire modes with $\sigma_j$ the corresponding eigenvalues and $C_j$ are the expansion coefficients obtained by a projection of the initial condition using the adjoint eigenmodes. 
After the wall-normal profile of $\tilde\eta_l^p$ is also expressed as an eigenmode expansion,
the normal vorticity becomes
\begin{equation}
  \hat{\eta} = \sum_j C_j \tilde{\eta}_j\ e^{-i\alpha \sigma_j t} +
  \sum_j D_{jl}\frac{e^{-i\alpha c_l t} - e^{-i\alpha \sigma_j t} }{
  \alpha c_l - \alpha \sigma_j} 
\label{eq:C5:eta-sol}
\end{equation}
where $D_{jl}$ are the expansion coefficients for the forcing term $U'\tilde{v}_l$.

\begin{figure}
\begin{center}
\includegraphics[width=0.6 \textwidth]{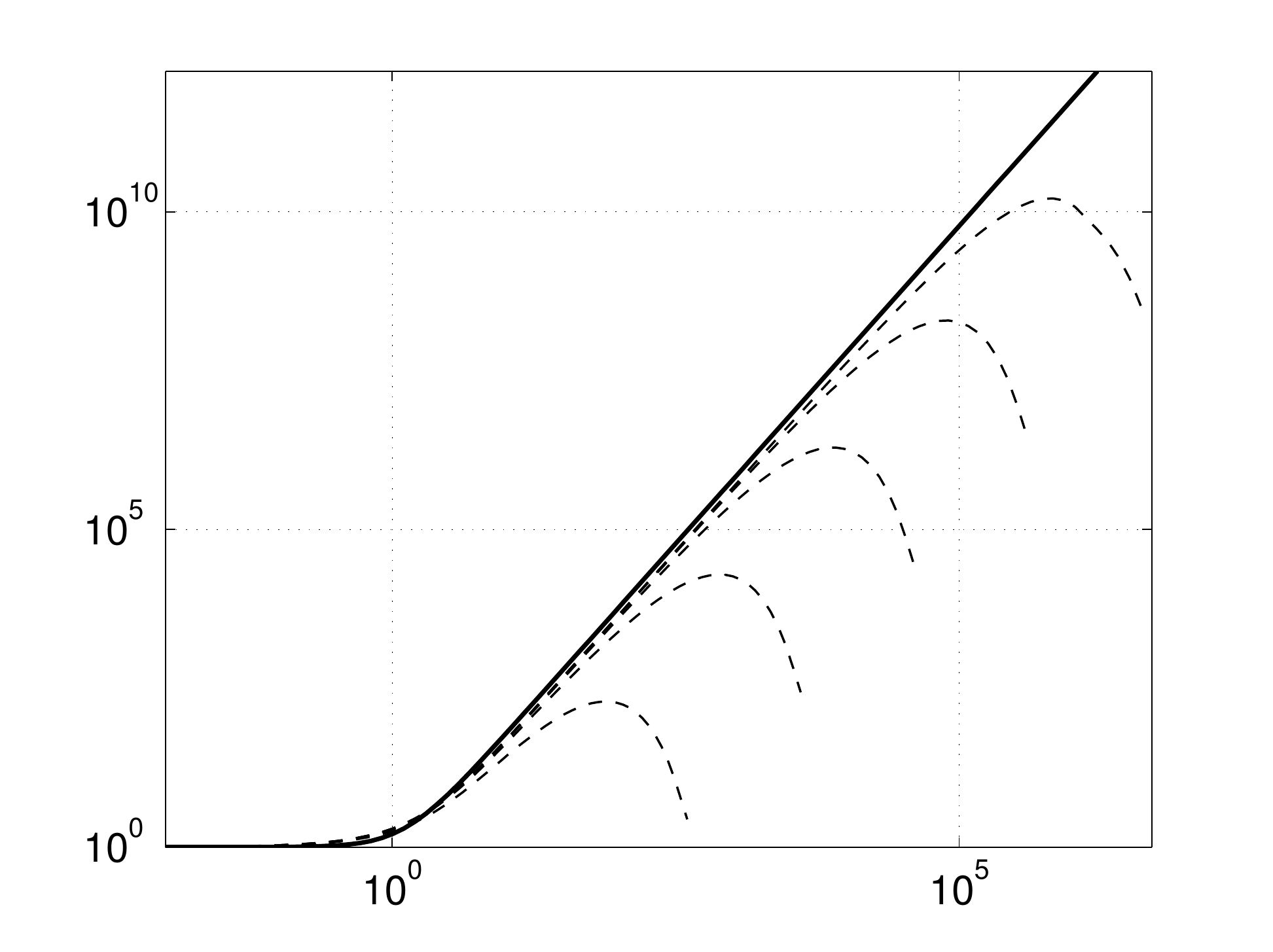} 
\put(-250,140){{$G(t)$}}
\put(-120,-1){{$t$}}
\end{center}
\caption{ Optimal transient growth, \lb{disturbance energy at the optimization time over the initial energy at time $t=0$}, for disturbances of wave vector $(\alpha,\beta)=(0,2)$ in the Poiseuille flow. Thin dashed lines indicate the growth for increasing values of the Reynolds number, $\Rey= 10^3, 10^4, 10^5, 10^6, 10^7$ and the thick solid line the inviscid limit, $G \approx 1 + 0.6 t^2$. }
\label{fig:limitRe}
\end{figure}

We now consider the evolution over finite times
in the limit
$\Rey \to \infty$ and $\alpha
\rightarrow 0$.
Multiplying the Orr-Sommerfeld and Squire equations with
their respective complex conjugates and integrating over the domain, it is possible to show that
\begin{align} 
\omega^{OS}_l &= \alpha c_l = -i\mu_l/\Rey \label{eq:C5:oneoverR1}, \\
\omega^{SQ}_j &= \alpha \sigma_j = -i\nu_j/\Rey \label{eq:C5:oneoverR2},
\end{align}
where $\mu_l$ and $\nu_j$ are positive quantities of order one.
In other words, the frequency of these modes is inversely
proportional to the Reynolds number and all tend to zero as the
Reynolds number approaches infinity. Substituting these into the relations for the expansion coefficients and 
Taylor
expanding for small $t/\Rey$ we find that
\begin{align}
  \hat{\eta} = \sum_j C_j \tilde{\eta}_j &\left[ 1 
                -\nu_j t/\Rey
  + {\mathcal{O}}\left(\frac{t^2}{ \Rey^2}\right)\right] \nonumber \\ 
  &- \sum_j
  iD_{jl}\tilde{\eta}_j t \left[1 - (\nu_l + \mu_j)\frac{t}{
  2 \Rey} + {\mathcal{O}}\left(\frac{t^2}{ \Rey^2}\right)\right].
\end{align}
Considering only term  of ${\mathcal O}(1)$ the solution to the Squire equation can be finally written as
\begin{equation}
  \hat{\eta} = \hat{\eta}_0 -i\beta U'\hat{v}_0t + {\mathcal{O}}
\left(\frac{t}{\Rey} \right).
\end{equation}
Again, as the Reynolds number approaches
infinity and for $\alpha =0$, the streamwise velocity perturbation (proportional to $\eta$ in this limit) grows linearly in time. At finite Reynolds numbers the growth is only transient.

The transient growth of streamwise elongated disturbances is clearly identified by an input-output or non-modal analysis \citep{Schmid2001,lucaAMR}. The stability analysis is casted as an initial value problem and the initial condition leading to the largest possible amplification over a finite time horizon is sought. Formally, the optimal growth is defined as
\begin{equation}\label{eq:Optimal_growth_definition}
G(t) \equiv \max_{q_0}\frac{||q||}{||q_0||}=\max_{q_0}\frac{||\exp(tL)q_0||}{||q_0||}=||exp(tL)||.
\end{equation}
The largest possible amplification is therefore the largest singular value of the evolution operator, $\mathcal{T}=\exp(tL)$, and the initial condition and the corresponding flow response are the left and right singular vectors, 
where for simplicity we re-write our linear system as $\frac{\partial q}{\partial t}= L q$. In equation (\ref{eq:Optimal_growth_definition}), a meaningful norm should be defined, typically the disturbance kinetic energy. 

\begin{figure}
\begin{center}
\includegraphics[width=0.45 \textwidth]{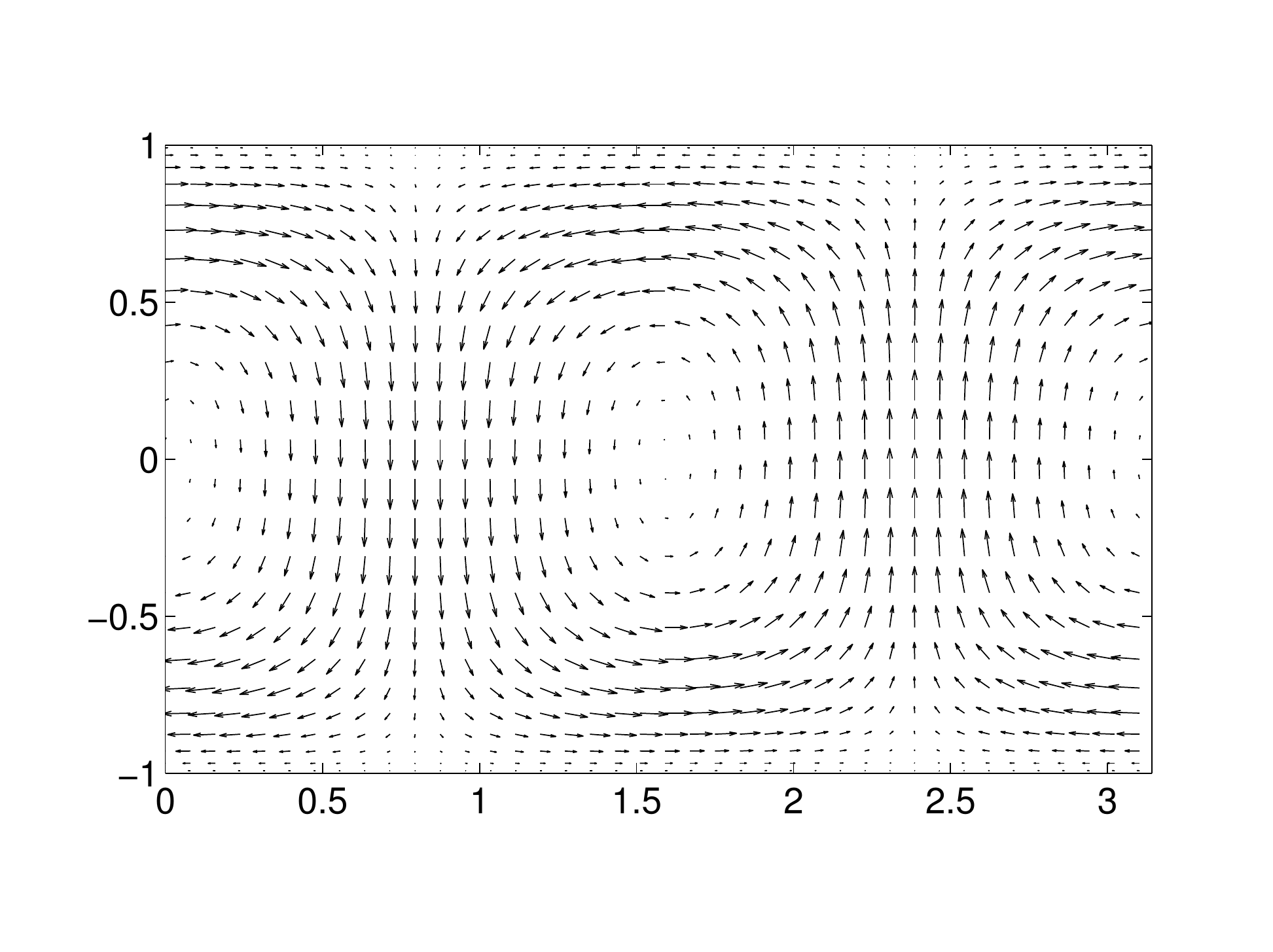} 
\includegraphics[width=0.45 \textwidth]{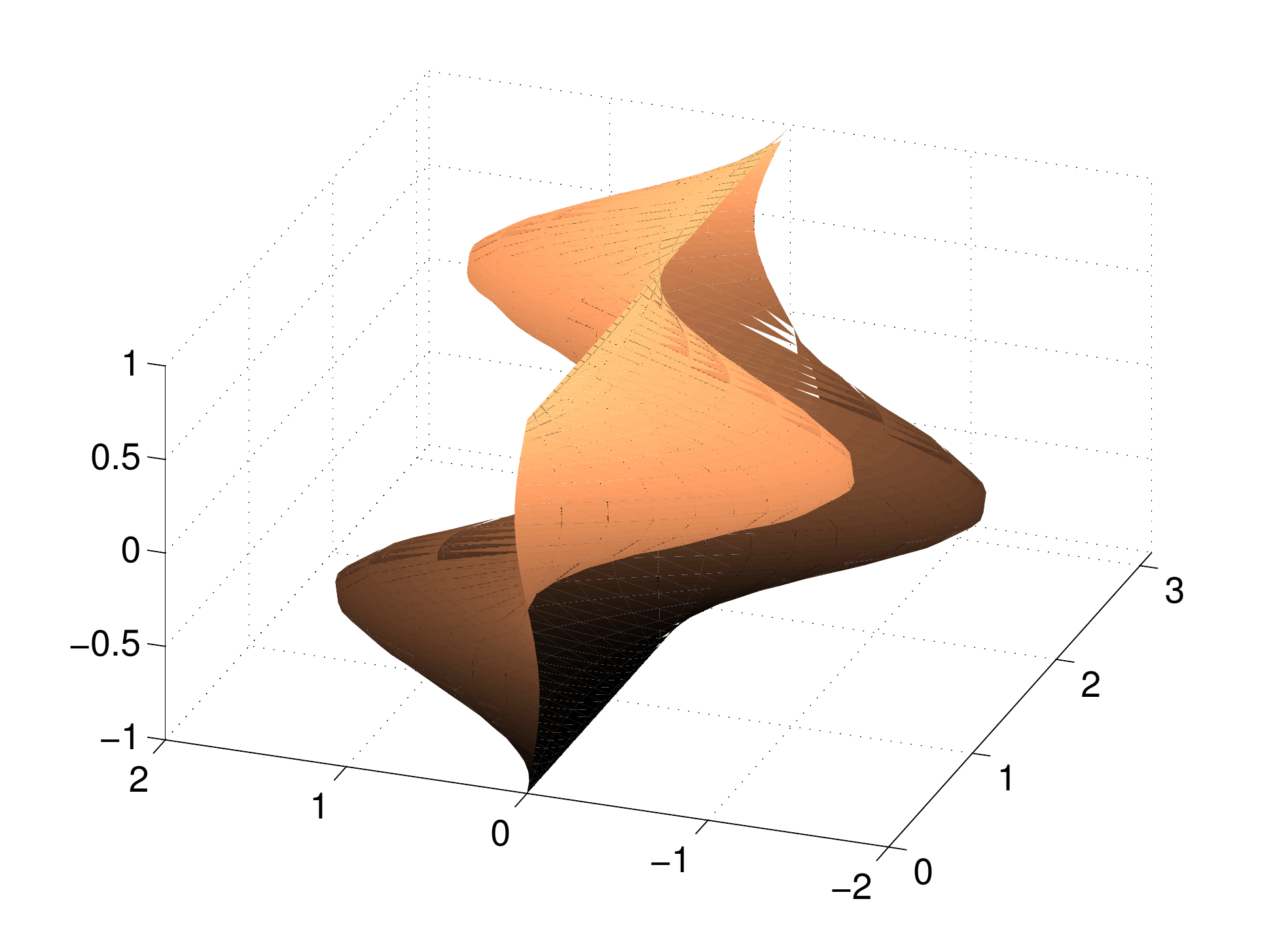} 
\put(-370,110){{$(a)$}}
\put(-360,70){{$y$}}
\put(-275,0){{$z$}}
\put(-180,110){{$(b)$}}
\put(-182,70){{$y$}}
\put(-20,25){{$z$}}
\put(-110,0){{$u$}}
\end{center}
\caption{ (a) Vector plot of the optimal initial condition for Poiseuille flow at $\Rey=2000$ and perturbations of wave vector $(\alpha,\beta)=(0,2)$ in the cross-stream $y-z$ plane. 
$(b)$ Streamwise velocity of the flow at the time of maximum growth $t=155$ with $G(t=155)=783$.}
\label{fig:optimal}
\end{figure}
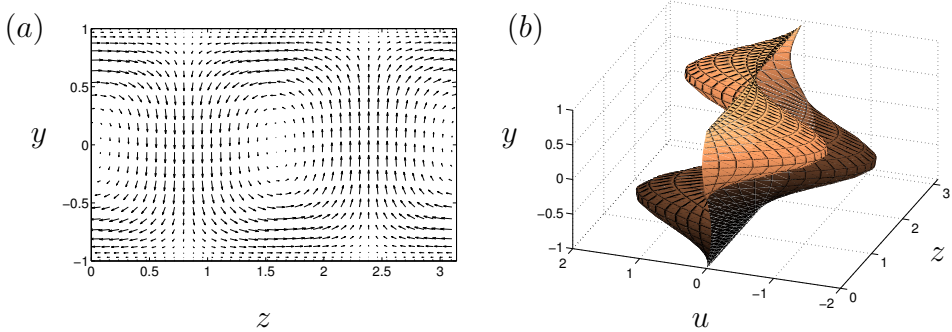

The flow receptivity can be analyzed by considering the flow response to a harmonic excitation. In this case, the optimal gain is given by the largest singular value of the resolvent operator $(\omega I - L)^{-1}$.
The largest transient growth in parallel shear flows is found for streamwise independent modes, $\alpha=0$. The energy growth is proportional to $\Rey^2$ while the time over which the growth extends scales as $\Rey$. The optimal transient growth versus time for different Reynolds numbers is displayed in figure~\ref{fig:limitRe}: the algebraic growth is clearly seen as the inviscid limit, as well as the scaling $t_{max} \propto \Rey$. The optimal initial condition and the corresponding flow at the time of maximum energy are depicted in figure~\ref{fig:optimal} for Poiseuille flow and the disturbance wave vector yielding the largest overall transient growth, $(\alpha,\beta)=(0,2)$. The initial condition consists of a pair of streamwise counter-rotating vortices extending across the channel width. The optimal response is two pairs of positive and negative streamwise streaks located on each half of the channel.

From a mathematical point of view, the non-modal growth can be explained by the non-normality of the
the linearised operator describing the flow dynamics and the
associated non-orthogonal set of eigenmodes \citep{reddyhenningson}.
If the state of the
system has a strong projection on some of these highly non-orthogonal
eigenmodes the energy of the flow can experience a significant
transient growth \citep{Schmid07}. This is now well-established, and indeed, any stability analysis should not consider only the long-time behavior dictated by the system eigenvalues but also the short-time flow response determined by a non-modal analysis \citep[see the tutorial by][]{lucaAMR}.

\subsection{Sensitivity to base flow modifications and role of linear amplification}

\begin{figure}
\begin{center}
\includegraphics[width=0.42 \textwidth]{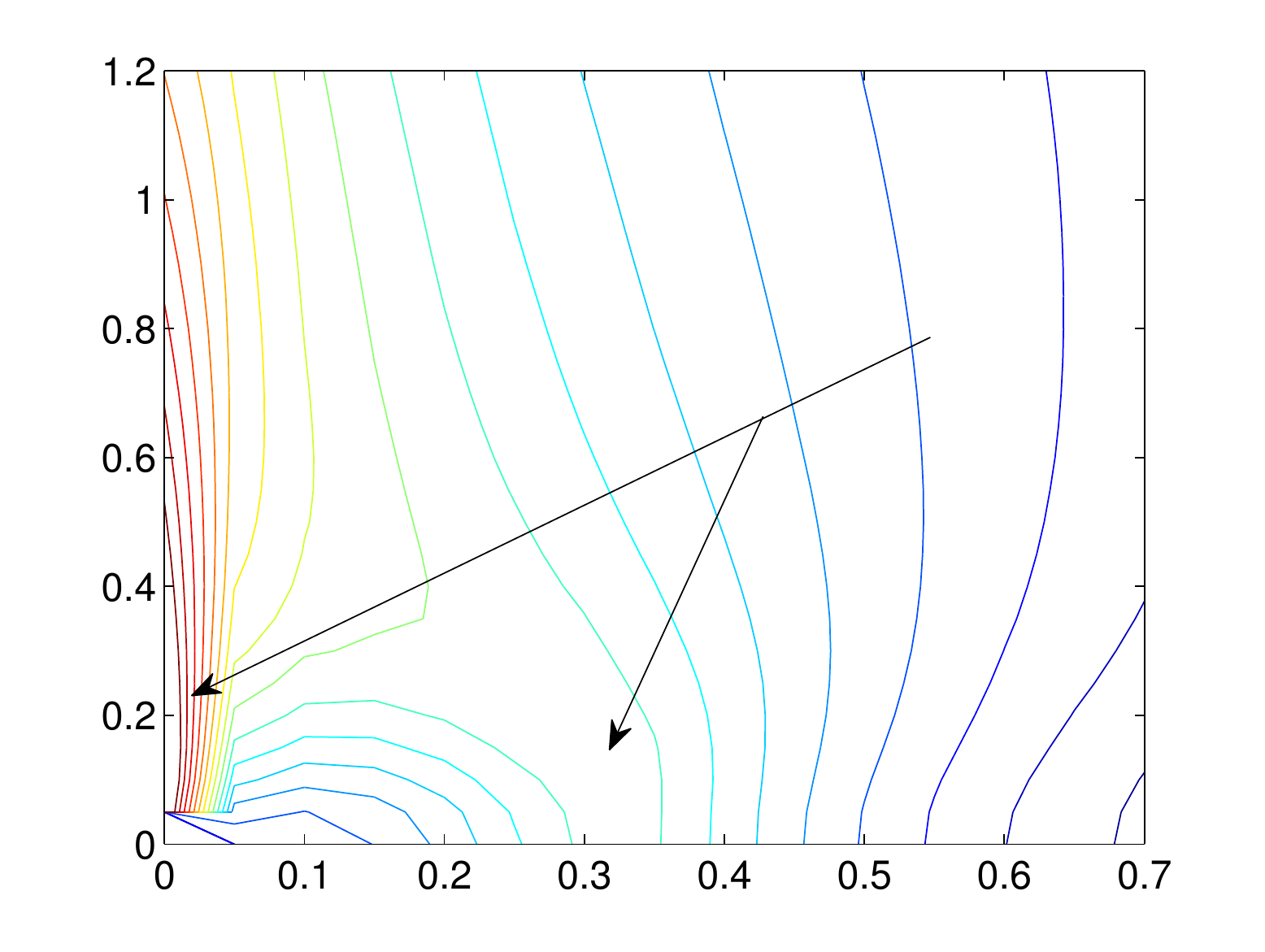} \hspace{1cm}
\includegraphics[width=0.42 \textwidth]{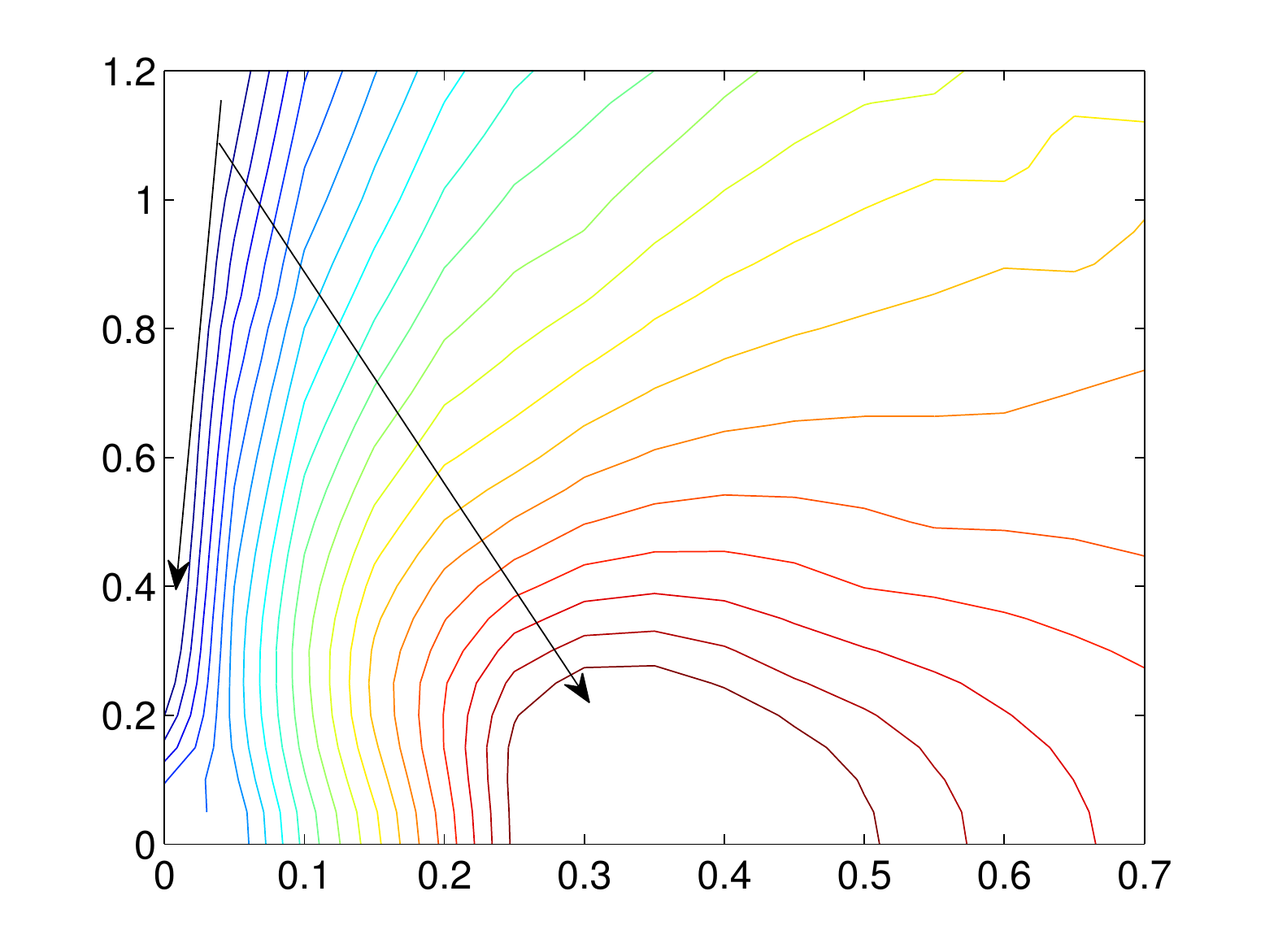} 
\put(-381,110){{\large $(a)$}}
\put(-181,110){{\large $(b)$}}
\put(-368,60){{$\beta$}}
\put(-287,-5){{$\alpha$}}
\put(-130,98){{$\frac{\mathbf{\nabla}_{U} \mathcal{R}}{\mathcal{R}}$ }}
\put(-170,60){{$\beta$}}
\put(-84,-5){{$\alpha$}}
\put(-290,70){{$\mathcal{R}$ }}
\end{center}
\caption{
$(a)$ Contour levels of maximum optimal response $\mathcal{R}$ in the $\alpha$--$\beta$ plane. Minimum level $10^{2}$, maximum level $10^{4.5}$, contour spacing $10^{0.15}$ (logarithmic). 
$(b)$ Contour levels of normalized sensitivity $\mathbf{\nabla}_{U} \mathcal{R}/\mathcal{R}$ of the optimal forcing with respect to base flow modification in the $\alpha$--$\beta$ plane for Blasius flow at $Re=400$. Minimum level $10^{0.7}$, maximum level $10^{2.6}$, contour spacing $10^{0.1}$ (logarithmic).}
\label{fig:sens_map}
\end{figure}
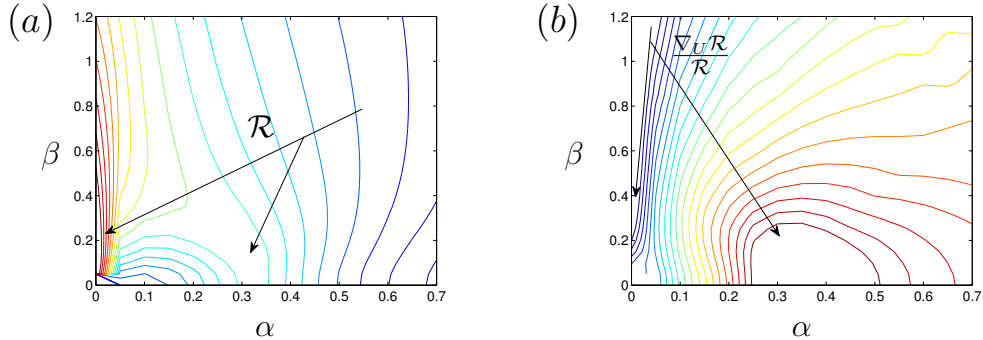

The lift-up effect turns out to be a very robust mechanism, ubiquitous in shear flows, often relevant if not dominating also in the presence of exponential instabilities. \lb{As also discussed below, the lift-up turns out to be dominant at moderate and high levels of external noise whereas modal instabilities, the so-called Tollmien-Schlichting waves, are responsible for transition in low-noise environments. 
Streaks are elongated structures modulated in the spanwise direction whereas the fastest growing Tollmieng-Schlichting waves are two-dimensional streamwise-dependent modes.}
The sensitivity of the lift-up effect to the presence of an external forcing and to non-homogenous boundary conditions (blowing and suction at the wall) is examined in \cite{brandt2011effect}. 
Using a variational technique, these authors derive an analytical expression for the gradient of the non-modal disturbance amplification with respect to base-flow modifications and show how it depends on the overlap between
the optimal initial condition (the streamwise vortices) and the flow at the optimization time (the streaks). When examining the flow receptivity to an external forcing in the frequency domain,  the sensitivity is similarly given by the overlap between 
the optimal forcing and the optimal response of the flow. 
As an application, the zero-pressure-gradient boundary
layer flow is examined where the different instability mechanisms of wall-bounded shear flows are at work. 
The analysis in \cite{brandt2011effect} is extended here to demonstrate the robustness of the lift-up mechanism: base-flow modifications can deeply alter the Tollmien-Schlichting instability
whereas the amplification of streamwise streaks is indeed a  strong process. 

We define $\mathcal{R}(\omega, \alpha,\beta; \Rey, U)$ the optimal response of the system to a time-periodic forcing of frequency $\omega$ and wave-number $\alpha$ and $\beta$, for a fixed Reynolds number $\Rey$ and base flow $U$, the Blasius boundary layer in this case.
Two peaks are clearly distinguishable when plotting the maximum amplification for the frequency $\omega$ maximizing the amplification $\mathcal{R}_\omega(\alpha,\beta;Re)$ in the $(\alpha,\beta)$ plane: the largest of them is located approximately at  
$(\alpha,\beta)=(0,0.2)$ and is due to the lift-up effect, while the second lower peak is for two-dimensional perturbations, $\beta=0$, with stream-wise wavenumber $\alpha\approx 0.3$, see figure\ref{fig:sens_map}(a).  
The sensitivity to variations of the base flow $\mathbf{\nabla}_{U} \mathcal{R}(y; \omega, \alpha,\beta)$, i.e.\ the gradient of the largest singular value of the resolvent operator with respect to base flow modifications, is shown in figure~\ref{fig:sens_map}(b). 
The wall-normal maximum of the gradient is chosen as a measure of the sensitivity.
To directly show the potential for stabilization, the gradient is normalized with the corresponding energy gain $\mathcal{R}_\omega(\alpha,\beta;Re)$.

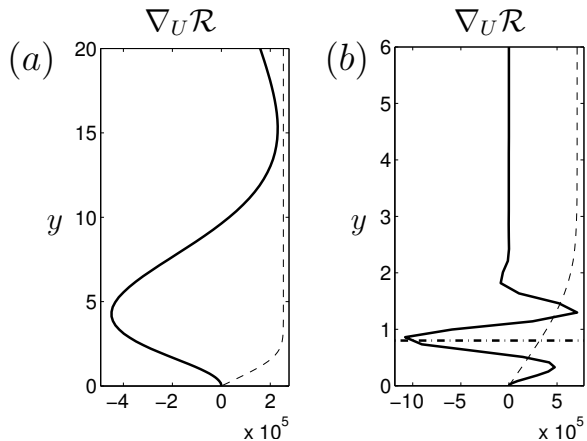
\begin{figure}
\begin{center}
\includegraphics[width=0.21 \textwidth]{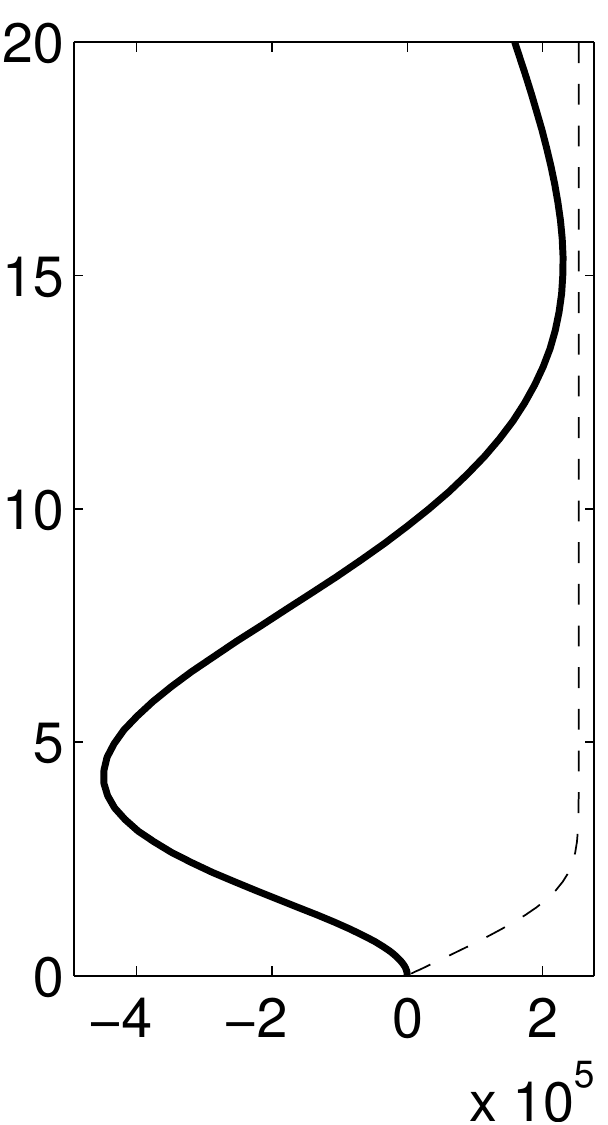}\hspace{1cm} 
\includegraphics[width=0.2 \textwidth]{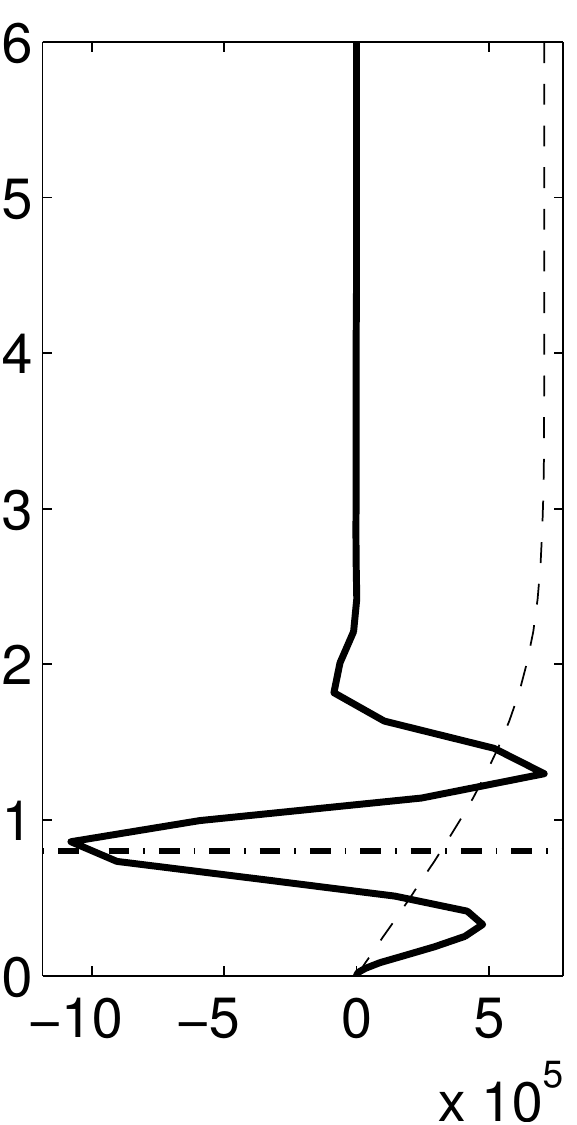}
\put(-220,140){{\large $(a)$}}
\put(-100,140){{\large $(b)$}}
\put(-205,80){{$y$}}
\put(-172,155){{ $\mathbf{\nabla}_{U} \mathcal{R}$}}
\put(-55,155){{ $\mathbf{\nabla}_{U} \mathcal{R}$}}
\put(-90,80){{$y$}}
\end{center}
\caption{Wall-normal profile of the gradient of the optimal response to base flow modification $\mathbf{\nabla}_{U} \mathcal{R}$ for parallel Blasius flow, $Re=400$
$(a)$  $\alpha=0$ and $\beta=0.2$;  $(b)$  $\alpha=0.3$ and $\beta=0$. The thin dashed line indicated the Blasius profile and the 
dashed-dotted line in $(b)$ indicates the location of the critical layer. See \cite{brandt2011effect}.}
\label{fig:grad}
\end{figure}

As shown in \cite{brandt2011effect}, the region where base flow modification may modify the disturbance amplification due to the lift-up effect spreads out all over the
flat plate and even upstream of it in the free stream, which make difficult to devise a passive effective control strategy. Conversely, modification in the shear layer, close to the wall, can significantly effect the exponential growth of the modal instabilities.
The wall-normal profiles of the gradient to base flow modifications $ \nabla_U \mathcal{R}(y) $ for the two dominant instabilities in the parallel Blasius flow are reported in figure~\ref{fig:grad}. The weak sensitivity of the lift-up process is directly seen by examining the expression for the gradient
of the resolvent norm and is related to the so-called component-wise non-normality \citep{Chomaz05}. 
This concentrates the optimal forcing and response on different components of
the velocity field, so that the overlap between the two perturbation vectors is minimal.

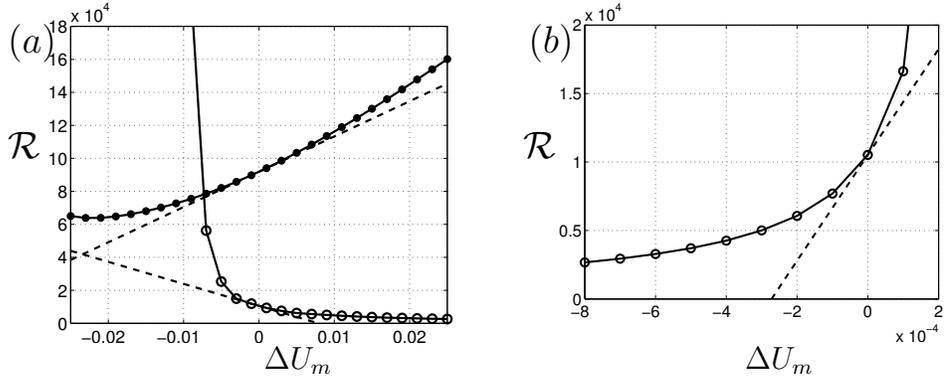
\begin{figure}
\begin{center}
\includegraphics[width=0.4 \textwidth]{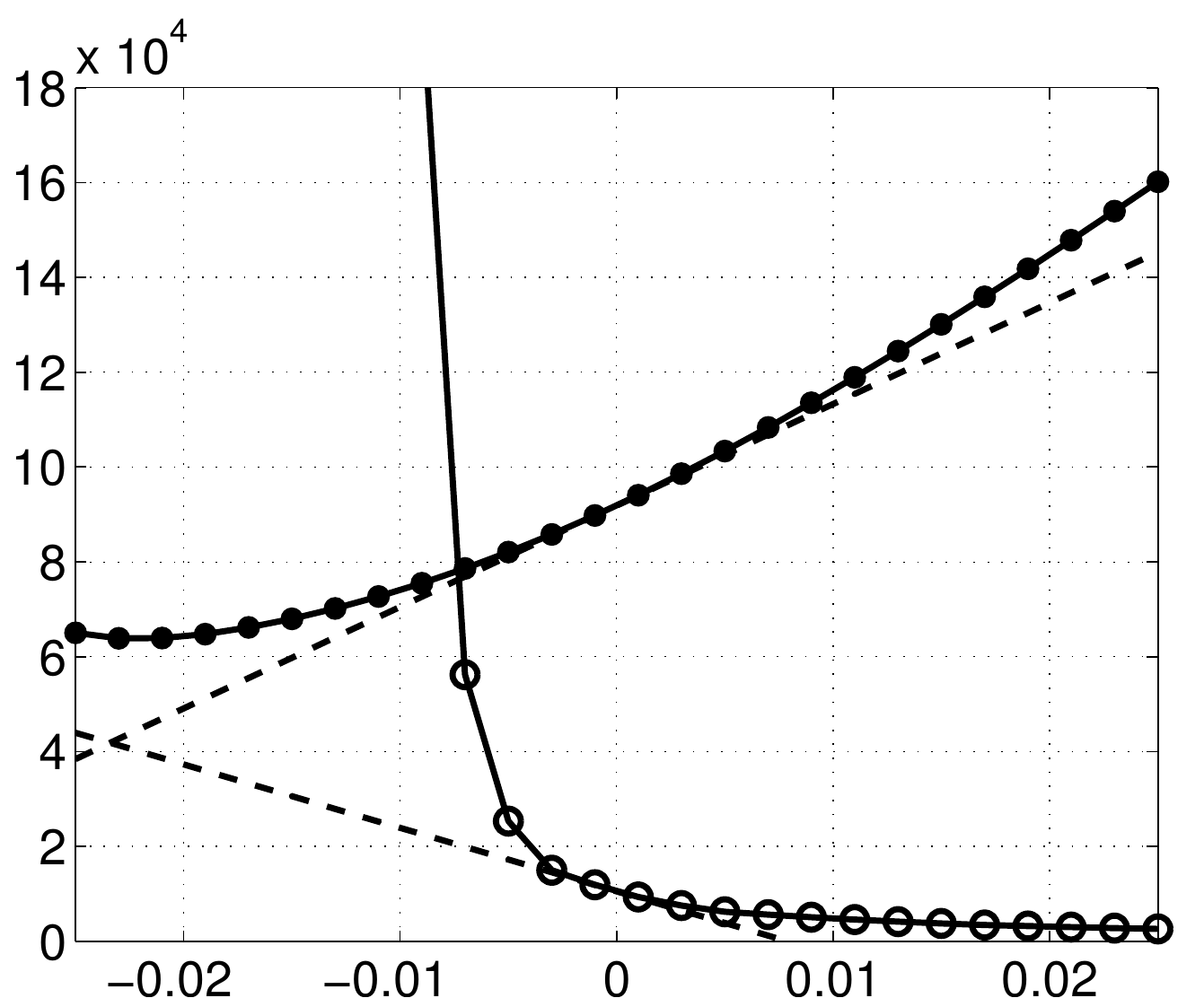} \hspace{1cm}
\includegraphics[width=0.37 \textwidth]{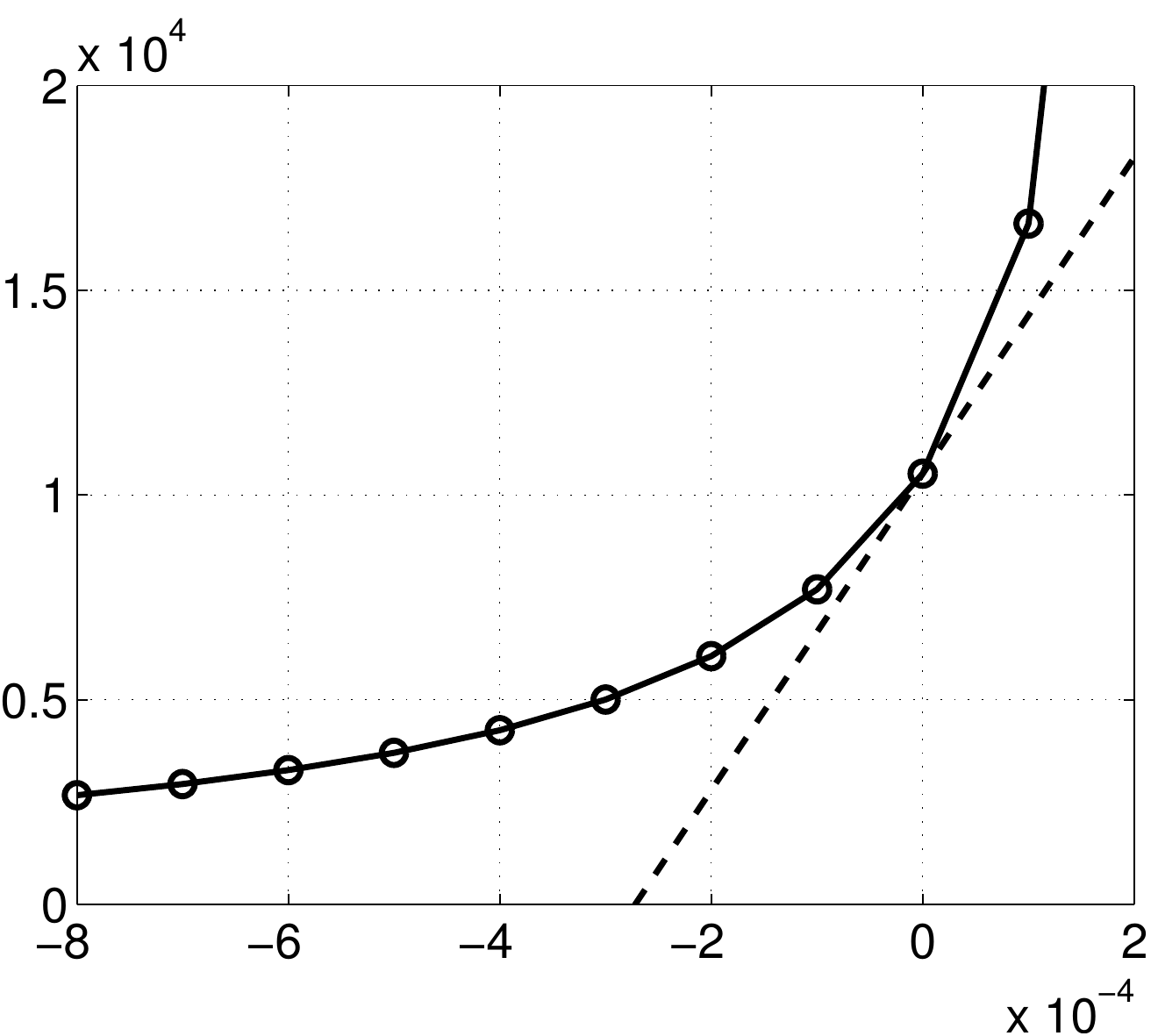}
\put(-355,110){{\large $(a)$}}
\put(-355,70){{\large $\mathcal{R}$}}
\put(-158,110){{\large $(b)$}}
\put(-160,70){{\large $\mathcal{R}$}}
\put(-262,-10){{ $\Delta U_m$}}
\put(-80,-10){{ $\Delta U_m$}}
\end{center}
\caption{Flow optimal response versus maximum of the wall-normal amplitude of the base flow variation $\Delta U_m$. 
Filled symbols are used for forcing with $\alpha=0$ and $\beta=0.2$, while open symbols for $\alpha=0.3$ and $\beta=0$; $Re=500$. The dashed line indicates the prediction with the local linear gradient.
$(a)$ The Blasius profile is modified by the gradient in figure~\ref{fig:grad}$(a)$; $(b)$ base flow modified by the gradient in figure~\ref{fig:grad}$(b)$.}
\label{fig:resp_grad}
\end{figure}

For the case of boundary-layer flows, where two distinct instability mechanisms are at work, it is relevant to examine how variations of the base flow that are beneficial to one type of disturbances can affect the other. To this aim, we consider modifications of the base flow which are optimal to reduce the lift-up or the Tollmien-Schlichting mechanism (as presented in figure~\ref{fig:grad}) and study the resulting flow behavior in the wave-number plane. $\Delta U_m=\max_y \left[ U(y)-U_B(y) \right]$ is used to measure the largest departure of the modified base flow $ U(y) $ from the Blasius profile $ U_B(y) $ and results are presented for $Re=500$ where most dramatic effects can be seen. 
Figure~\ref{fig:resp_grad}$(a)$ depicts the variations of the maximum response $\mathcal{R}_\omega$ when using the profile in figure~\ref{fig:grad}$(a)$ targeting the streaks to perturb the base flow. The filled symbols show how the response at $\alpha=0, \beta=0.2$ varies, open symbols the behavior of two-dimensional waves with $\alpha=0.3$ and $\beta=0$, while the dashed line indicates result one would obtain extrapolating the gradient derived above.
 When considering perturbations characterized by $\alpha=0.3, \beta=0$, however, one can notice that base flow modifications, positive for streak amplifications, are now detrimental to Tollmien-Schlichting waves and vice versa. At $Re=500$, departures of about 0.5\% induces a modal instability of the Tollmien-Schlichting type. 
This effect is significantly reduced at lower Reynolds numbers: for example at $ Re=300 $, the flow does not become unstable for distortions of the order of 2-3\% and the most dangerous external disturbances are streaks. The results suggest that large base flow modifications are needed to significantly affect the nonmodal lift-up effect, a strong limitation coming from the fact that the flow may become susceptible to time-dependent instabilities.
The results in the figure also suggest the range of validity of predictions based on the gradient $\mathbf{\nabla}_{U} \mathcal{R}$ computed using the unperturbed Blasius profile, in other words the validity of the linear approximation. Modifications of the order of 1\% of the free-stream velocity can be well captured by the linear model while, for larger $\Delta U_m$, we observe a reduction of the stabilizing effect at negative values and an increase of the destabilization at positive distortions.

Figure~\ref{fig:resp_grad}$(b)$ reports the case of modifications of the base flow which are optimal to reduce the amplification of
Tollmien-Schlichting waves. Data are not shown for the case of disturbances with $\alpha=0, \beta=0.2$ since the base flow modification amplitude considered here do not induce any significant variation in $\mathcal{R}_\omega$. Very minute distortion of wrong shape can induce soon an unstable flow whereas significant reduction can be achieved with the correct modifications. This indicates that it may be possible to successfully target Tollmien-Schlichting waves instability and thus stabilize the flow in cases where non-modal effects are of less importance, e.g.\ boundary layers on wings with low levels of ambient vortical disturbances where acoustic waves and roughness trigger a transition scenario dominated by local convective instabilities. 
Note finally that the values of $\mathcal{R}_\omega$ reported are obtained for the same frequency as in the case of Blasius flow since it has been verified that the frequency of maximum response is not changed when altering the base flow. 
In summary, our results show that variation of the base flow reducing the streak amplification can easily lead to
more unstable Tollmien-Schlichting waves. Weak variations of the shear layer close to
the wall can largely affect the Tollmien-Schlichting amplification while have no effect on the lift-up effect.

\subsection{Subcritical transition in shear flows}

The non-modal amplification of streamwise streaks is thus identified as the first disturbance growth mechanism in the case of subcritical transition in shear flows. The transition to turbulence triggered by a localized disturbance in plane channel flow and the role of streamwise elongated modes was documented in \cite{henningson1993mechanism}. Several later studies considered the role of streak growth in channel flow transition \citep{sreddy:1998} as a first linear stage in the transition process.

More recently, using a nonlinear optimisation technique in a periodic computational domain, few groups have identified the
perturbations in canonical shear flows transitioning with the least initial kinetic energy, the so-called minimal seed \citep{Pringle2010,monokrousos2011nonequilibrium}.
These appear to be  spatially localised perturbations and the following route to turbulence display several known linear mechanisms 
one after the other: Orr mechanism \citep{orr:1907}, lift-up, streak bending and
breakdown \citep{duguet2010towards,duguet2013minimal}.
In \cite{CherubiniPRE2010}, non-linear optimal perturbations have been computed for a Blasius boundary-layer flow. 
The results show that non-linear optimal perturbations leading to transition 
are formed by vortices inclined in the streamwise direction surrounding a region of intense streamwise disturbance velocity.  This minimal seed grows very rapidly in time due to the transport of the base flow momentum by the disturbance (lift-up) along the inclined vortices, resulting in an amplification of the streamwise velocity disturbance, which is then dislocated by the initial vortices due to self-interaction of the perturbation with itself. This mechanism has been called ''modified lift-up'' in \cite{CherubiniJFM2011} since it does not create streaks but $\Lambda$-shaped structures of slow and fast fluids.
In the last stages of transition, the redistribution of the vorticity due to the non-linear mixing induces the creation and the lift-up of a spanwise vorticity zone (the arch vortex) connecting the two initial neighboring vortex structures, constituting a main hairpin vortex, which then releases new vortical structures due to inflectional instabilities related to the base flow modifications. A similar mechanism of energy growth linked to a "modified lift-up" have been also found for nonlinear optimal perturbation in the Couette flow \citep{CherubiniJFM2013}.

\subsubsection{Bypass transition}

\begin{figure}[!t]
 \begin{center}
\includegraphics[width=.8\linewidth]{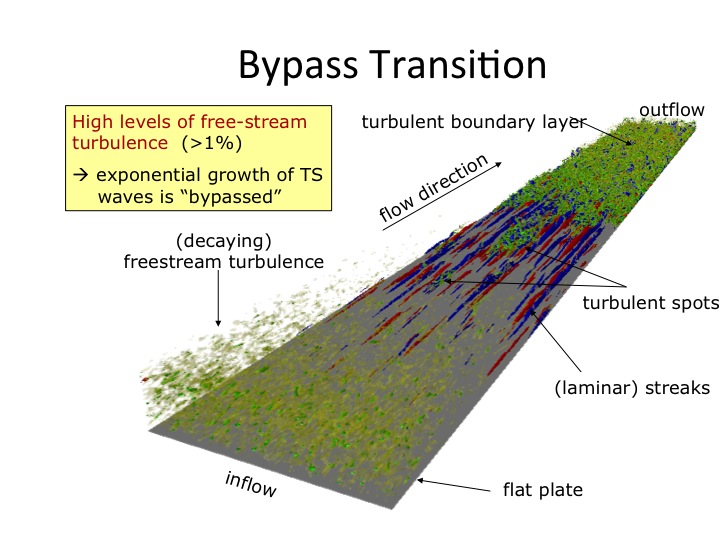}
\end{center}
\caption{Sketch of bypass transition in a boundary layer exposed to high levels of free-stream turbulence. Details of the simulations are reported in \cite{luca_phil,Schlatter08}. Visualization courtesy of Philipp Schlatter.}\label{fig:bypass}
\end{figure}

We demonstrate the importance of the lift-up mechanism in the case of the laminar-turbulent transition in a zero-pressure-gradient boundary layer subject to high levels of free-stream turbulence, see figure~\ref{fig:bypass}. 
The visualization of the transition under free-stream
turbulence is extracted from the numerical simulations presented in \cite{luca_phil,Schlatter08} \lb{at supercritical conditions}.
Streamwise streaks can be seen to form close to
the computational inlet, followed by streaks oscillations and
turbulent spots until the flow eventually becomes fully turbulent. 
Such a scenario is usually referred to as bypass since the transition occurs bypassing the exponential growth of the
Tollmien-Schlichting waves that would have been observed in low-noise environments. 
This is one  relevant application where significant
energy growth can be observed when the flow is asymptotically linearly stable and where a disturbance other than that linearly unstable is found to trigger transition also at supercritical conditions: in this case streamwise \lb{elongated} streaks induced by the lift-up effect that dominate over the slow viscous growth of the two-dimensional Tollmien-Schlichting waves. Experiments on bypass transition first clearly revealed the role of streaks \citep{WestinBoiko1994,BoikoWestin1994,MatsuAlf2001,Lundell03}. From theoretical analyses,
it is shown that  the upstream
perturbations which undergo the largest possible downstream growth consist of
streamwise counter-rotating vortex pairs at the plate leading edge, see
\cite{andersson99,luchini}.  The same process of amplification of low-frequency disturbances is identified also considering periodic vortical disturbances in the incoming free stream \citep{schrader2010receptivity,schrader2012nonlinear}.

After the primary energy growth due to the lift-up effect, the flow is
in a more complicated laminar state where strong nonlinear
interactions can come into play, cf. figure~\ref{fig:bypass}. As the streaks grow in strength, they become susceptible
to high-frequency secondary instabilities due to the presence of both
wall-normal and spanwise inflectional velocity profiles
\citep{brandt02,lucaEJMB}. These secondary instabilities manifest
themselves in symmetric and antisymmetric streak oscillations, which
are precursors to the formation of localised regions of chaotic swirly
motion, the so-called turbulent spots
\citep{Zaki2005,Schlatter08,mans07}, typically triggered inside the boundary layer by streak interactions and instabilities \citep[see e.g. the recent simulations and experiments in][]{BrandtLange2008,nolan2010quadrant,nolan2012particle}.  The leading edge of
a spot travels at about the free-stream velocity $U_\infty$ while the
trailing edge at half this speed. The spots become therefore more
elongated and eventually merge: a fully-developed turbulent boundary
layer is observed. 

The bypass transition scenario is observed when the boundary layer is
subject to free-stream turbulence levels higher than 0.5-1\%
\citep{Fransson2005}.  As described below, the flow reproduces, though on
a larger scale, the near-wall dynamics of wall-bounded turbulence, see
e.g. \cite{robinson91}, and it is therefore and ideal test configuration for a better understanding and 
possible control of turbulent flows \citep{lundell2007reactive,Monokrousos2008}.

As shown in the figure, in boundary layer exposed to free-stream turbulence, low-frequency disturbance enter the shear layer and then amplify in the form of streaks. The ability of the low-frequency modes to penetrate the boundary layer and the filtering of high frequency disturbances is refereed to as shear sheltering effect.  This was investigated by \cite{jacobs98} and \cite{zaki2009shear}, see also the recent review by \cite{zaki2013streaks}, and it is a precursor of the lift-up in boundary layers. The physical interpretation of the sheltering behaviour of the
shear is given in Zaki and Saha and briefly reported here. The filtering is determined by the ratio of the timescale associated with the convection of a wave relative to an observer moving at a lower velocity inside the shear and the timescale of the wall-normal diffusion into the boundary layer. In the limit of weak shear (small relative velocity) or very long waves, an observer inside the shear layer is reached by the diffusion of the disturbances in the free stream as the observer does not see significant variations of the external disturbances while traveling  downstream. Conversely, in the limit of strong shear (large relative velocity) or short waves, 
the observer will see several waves passing by and the net average effect at the observer location will be small.
As summarized in \cite{zaki2013streaks}, when the diffusion time is relatively short (long waves), an
observer inside the shear layer can ÒresolveÓ the free-stream disturbance that is convected at a relatively
higher speed. Under strong shear, or when the diffusion time is relatively long (short waves), the
observer cannot ÒresolveÓ the free-stream disturbance.
Note that disturbances can also be advected directly inside the shear layer at the leading edge of, e.g.\ a compressor or turbine blade \citep{zaki2010direct}, as studied numerically in \cite{schrader2010receptivity,schrader2012nonlinear}. The combination of forcing form the free stream and inflow disturbances was discussed in \cite{WestinBoiko1994}.

\section{Turbulent flows}

\subsection{Turbulent streaks}

Streamwise vortices and streaks are also fundamental structures in the near-wall
region of turbulent shear flows where the vortices seem to be directly or indirectly related to streak
instabilities. In fact, the structures identified in bypass transition show a close resemblance to the ones
detected in turbulent wall flows.
\cite{kimkrey} were among the first to clearly show the importance of
local intermittent inflectional instability riding on near-wall streaks in the bursting events, those associated
with periods of strong turbulent production. These authors observed three oscillatory types of
motion of the streaks: a growing streamwise vortex, a transverse vortex and a wavy
motion in the spanwise and wall-normal directions. Later experiments by \cite{swearingen:87} compared and related the latter two modes observed in \cite{kimkrey} to the secondary varicose and sinuous instability of streamwise vortices measured during
transition on a concave wall.

\begin{figure}[!t]
 \begin{center}
\includegraphics[width=.6\linewidth]{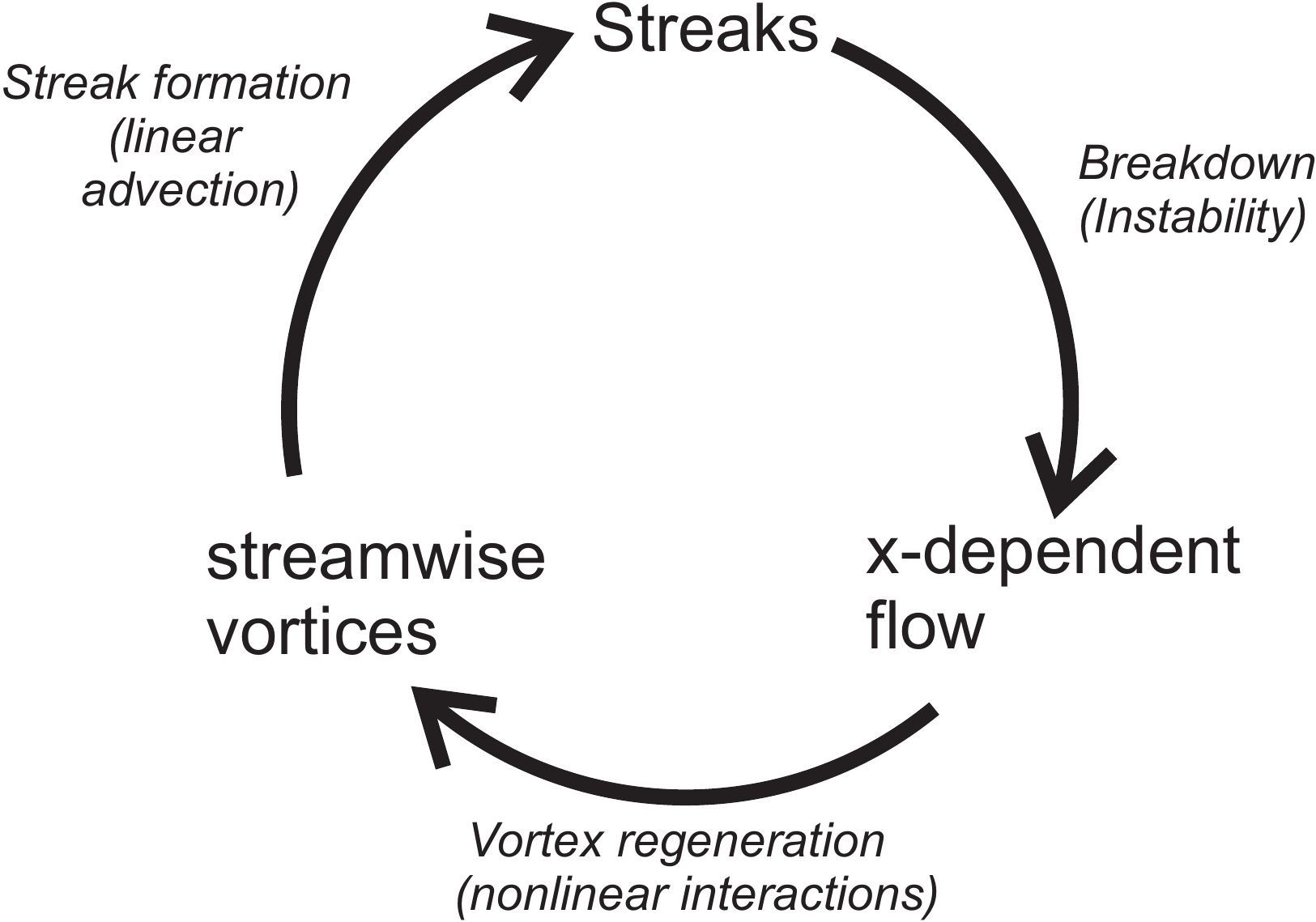}
\end{center}\caption{Sketch of the regeneration cycle of wall turbulence, see \cite{hamkimwal95, waleffe:97}.}\label{fig:SSP}
\end{figure}

As a consequence of these observations, the linear amplification of streamwise elongated structures, the lift-up or algebraic growth, is a key ingredient in any reduced-order model of turbulence aiming to reproduce self-sustaining oscillations with a system of ordinary differential equations. 
The simplest regeneration cycle proposed to explain the basic dynamics of wall-bounded turbulent flows is probably the one by \cite{hamkimwal95, waleffe:97}, see figure~\ref{fig:SSP}. This consists of three steps: i) generation of streaks induced by streamwise vortices (the linear lift-up effect), ii) streak breakdown via inflectional secondary instabilities (bursting events), iii) regeneration of elongated vortices by nonlinear interactions between oblique modes originating at the streak breakdown. More complicated models, based on Galerkin projection, also always need a linear amplification mechanism for transition and sustained turbulence, the lift-up effect \citep[e.g.][]{moehlis2004low}.

Recently, the computation of optimal energy amplifications has been extended to turbulent flows. 
Early attempts have consisted in using the turbulent mean flow profile in the Orr-Sommerfeld equations while neglecting  the Reynolds stresses associated to turbulent fluctuations \citep{Butler1993}. 
Following this approach in e.g. turbulent channel flows it is found that the perturbations leading to the maximum growth are streamwise uniform and spanwise periodic with wavelength about $3\,h$, where $h$ is the channel half-width, which is almost the same value selected in the laminar case. The typical spanwise spacing of one hundred wall units, characteristic of the near wall region streaks, is obtained by constraining the optimization to times of the order of the eddy turnover time \citep{Butler1993}.
Progress has been made by using the eddy viscosity $\nu_T(y)$ associated with the turbulent mean flow to model turbulent Reynolds stresses in the spirit of early modal linear investigations of turbulent channel flows, see \cite{Reynolds1972}. 
This results in Orr-Sommerfeld-Squire operators generalized to include the effect of a non-uniform viscosity.
The first incorrect expression of these operators \citep{delAlamo2006} has been amended in later studies \citep{Cossu2009,Pujals2009} to give:
\begin{eqnarray}
\label{eq:OSSQop}
\mathcal{L_{OS}}&=&-i\alpha(U \Delta-U'')
       +\nu_T \Delta^2+2 \nu_T' \Delta {\mathcal D} +\nu_T'' \Delta,\\
\mathcal{L_{SQ}}&=&-i\alpha U + \nu_T \Delta+\nu_T' {\mathcal D},
\end{eqnarray}
where 
$\Delta={\mathcal D}^2-k^2$.
Using this approach, without any further restriction on the optimization times, two peaks for the transient energy growth are found for $\alpha=0$. 
The main peak scales on external units with an optimal spanwise wavelength $\lambda_z \sim 4\, h$,
in fair agreement with the spanwise spacing of large-scale streaky motions in the outer region;  the associated maximum energy growth increases proportionally to a Reynolds number based on the outer units $Re_{out}=U_c h / \nu_{T,max}$   \citep{Pujals2009}.  
The secondary peak, independent of the Reynolds number, scales in inner (wall) unit and corresponds to $\lambda_z^+ \approx 90$, i.e. the most probable spanwise wavelength of near wall streaks \citep{kline67,Smith1983}. 
Optimal streaks corresponding to this secondary peak correspond well to the observed near wall-streaks.
Structures with scales broadly lying between these two peaks correspond to log-layer streaks.
If, instead of the optimal temporal growth, the optimal response to stochastic or harmonic forcing are considered, the same double-peaked amplification curve is obtained when pre-multiplied respectively by the spanwise wavenumber and its square \citep{Hwang2010c}.
Similar results are found in the turbulent Couette flow \citep{Hwang2010}, pipe flow  \citep{Willis2010} and boundary layer \citep{Cossu2009} where the spatial transient growth of coherent streaks has also been  observed experimentally
\citep{Pujals2010b}.

These recent studies confirm the relevance of the lift-up effect to explain the presence of streaks in turbulent flows that was anticipated by \cite{landahl90}. The coherent streamwise structures can efficiently extract energy from the mean flow via a coherent lift-up effect. This mechanism is predicted to be potentially active for spanwise scales ranging from those of the near-wall streaks (roughly one hundred wall units) to those of the large scale motions. This has been confirmed in more recent investigations \citep{Hwang2010b,Hwang2011} where it has been shown that self-sustained turbulent processes are active at all scales in turbulent channel flow without requiring energy input from either smaller or larger scales.  Furthermore, the coherent lift-up has been used to enforce passive control e.g. to reduce turbulent drag in pipe flow \citep{Willis2010} or to suppress turbulent separation on 3D bluff bodies \citep{Pujals2010}.

A similar linear model-based description of the scaling and location of turbulent fluctuations
in turbulent pipe flow is presented in \cite{mckeon1} and used to understand the behaviour of
the very large scale motions. The model is derived by treating the nonlinearity in the
perturbation equation (involving the Reynolds stress) as an unknown forcing, yielding
a linear relationship between the velocity field response and this forcing. 
This formulation of the Navier--Stokes equations
is designed to examine the receptivity of turbulent flows to forcing, see also \cite{jovbamJFM05}.
A singular value decomposition of the resolvent identifies the forcing shape that will lead to the largest velocity response at a given wavenumber--frequency
combination, as in the receptivity to forcing discussed above.
This approach is able to predict packets of hairpin vortices and other structures in turbulence under the assumption of a turbulent mean flow \citep{mckeon2}, demonstrating once more the importance of linear mechanisms such that identified by Ellingsen and Palm in turbulent shear flows. 

In this context, it is interesting to note the work of \cite{Gayme2011}. These authors study the input-output response of a streamwise constant projection of the Navier-Stokes equations for plane Couette flow.
The results of their analysis agree with previous studies of the linearized Navier-Stokes equations where the optimal energy amplification corresponds to minimal nonlinear coupling. On the other hand, the model provides evidence that the nonlinear coupling is responsible for creating the deformation of the turbulent velocity profile. This indicates that there is an important tradeoff between energy amplification, which can only be induced by a linear mechanism, and the seemingly nonlinear momentum transfer that produces a turbulent-like mean profile.


\subsection{The regeneration of vortical structures and the lift-up mechanism}

A full statistical analysis of turbulence production and its connection to the lift-up effect has been performed by the group in Rome \citep{gualtieri2002scaling} by considering a homogenous shear flow. It is reported here as the fully turbulent counterpart of the phenomenological models discussed above. Indeed phenomenological models of streak formation and  breakdown, similar to those observed in transitional flows, can be found in  \cite{kaw,Schoppa-Hussain97,Jimenez-Pinelli,Schoppa02,jimenez-simens} and will not be further discussed here.

In the statistically steady state the homogeneous shear flow is characterized by 
large fluctuations of the turbulent kinetic energy, 
see e.g. \cite{pumir1996turbulence}.
Such fluctuations can be readily understood and explained in terms of the corresponding
regeneration cycle of vortical structure and the associated lift-up mechanism.
See also the discussion reported in the early studies by 
\cite{rogers1987structure,lee1990structure,kida1994dynamics,waleffe:97}
and the recent book by \cite{sagaut2008homogeneous}.

\begin{figure}
\centerline{
\includegraphics[scale=0.40]{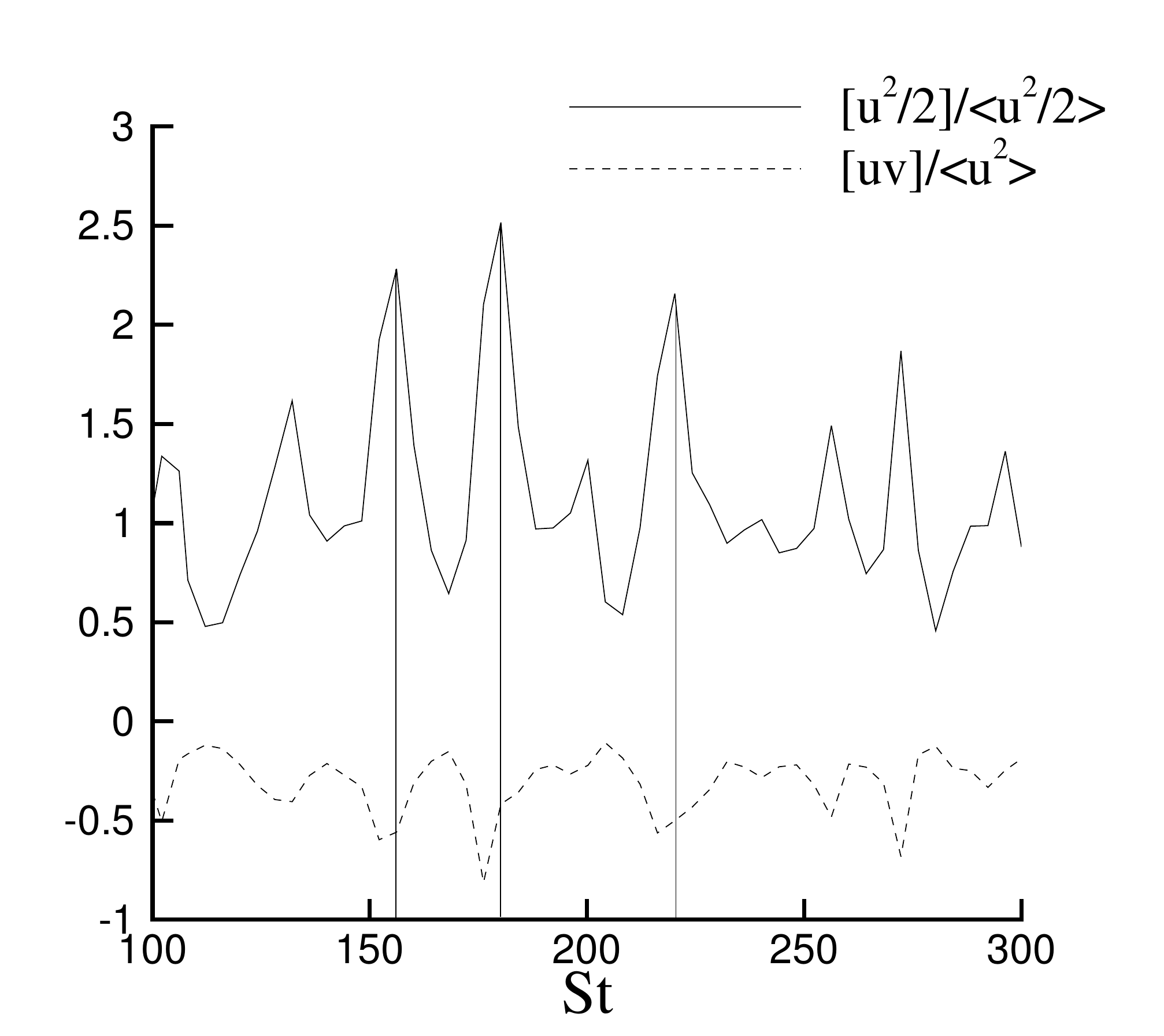}}
\caption{Time history of the spatially averaged turbulent kinetic energy (solid line) 
and Reynolds stresses with the sign changed (dotted line). \lb{Time is made non-dimensional with the mean flow shear S.} See \cite{gualtieri2002scaling}.
\label{fig:energy_snapshot} }
\end{figure}

\begin{figure}
\includegraphics[scale=0.32]{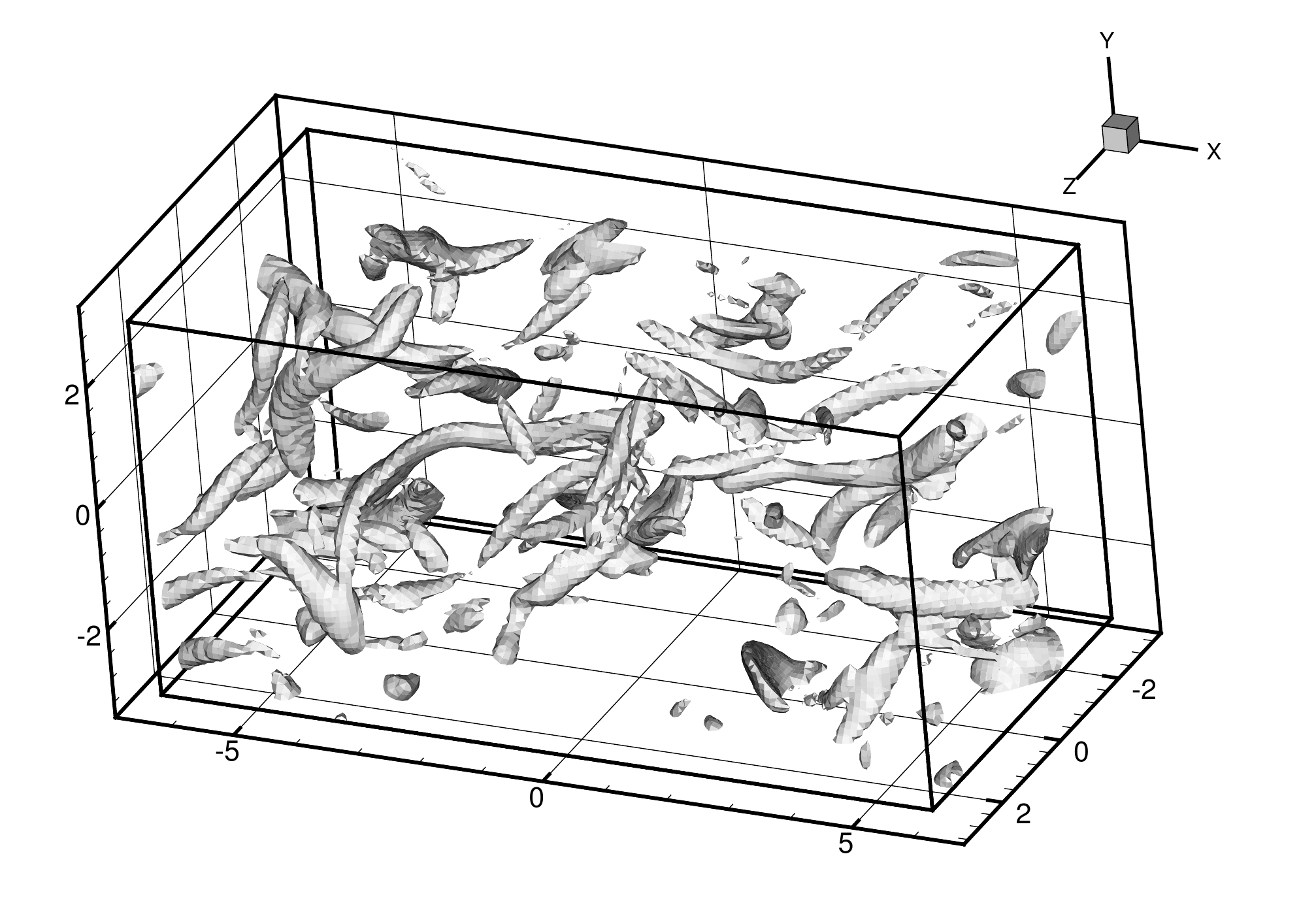}
\includegraphics[scale=0.32]{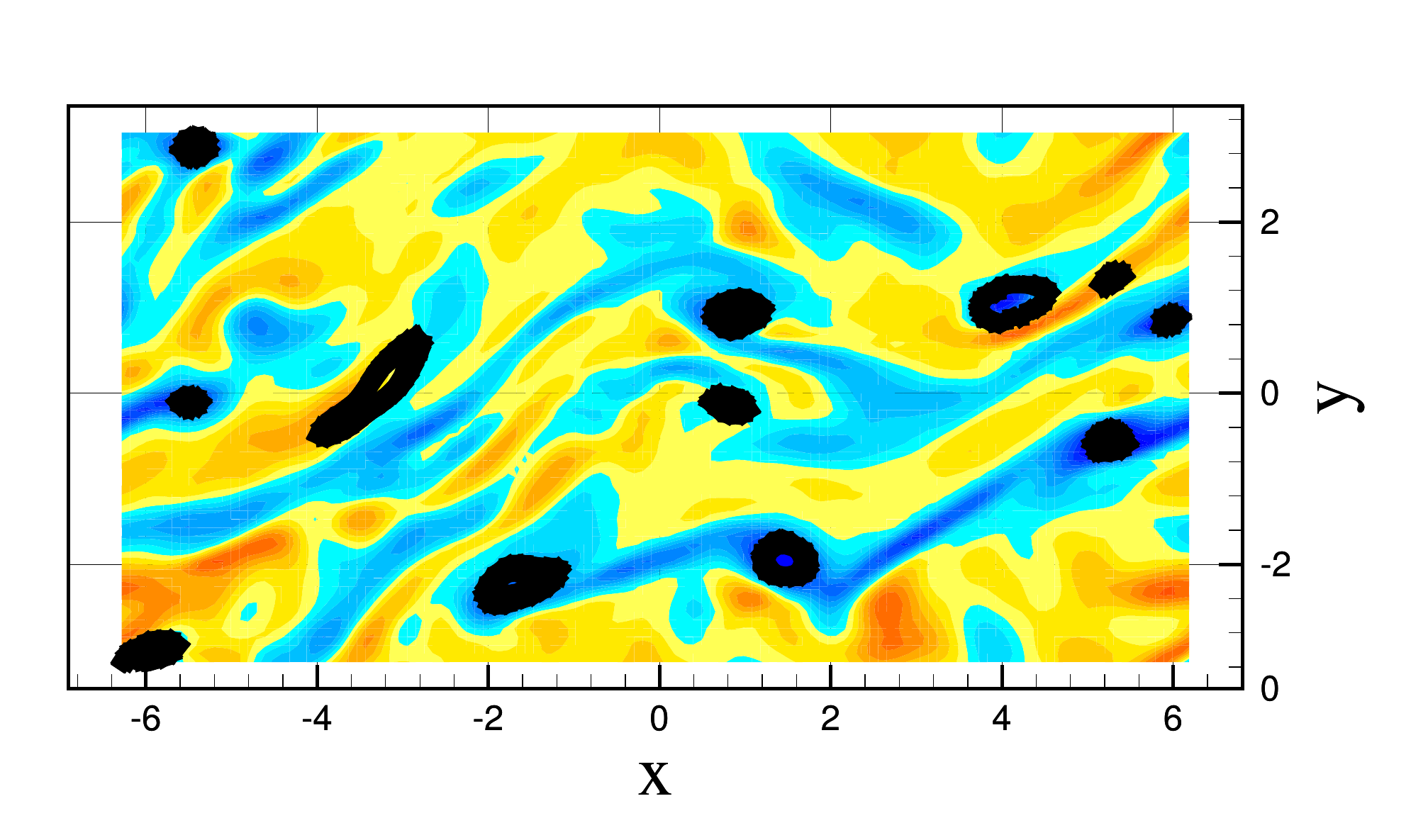}
\put(-380,95){{\large $(a)$}}
\put(-199,95){{\large $(b)$}}
\caption{$(a)$ Snapshot of the quasi-streamwise vortices and $(b)$ the associated vortex
sheets during the early stages of the
turbulent kinetic energy growth in a homogeneous shear flow  \citep[see also][]{gualtieri2002scaling}.
\label{fig:iso_D_growth} }
\end{figure}

Figure~\ref{fig:energy_snapshot} reports a typical history of the spatial average 
turbulent kinetic energy and highlights how the energy bursts 
are correlated to large negative values of the Reynolds shear stresses.
The data are obtained for the simple case of homogeneous shear flow.
The bursts are induced by large energy injections from the mean flow 
to the velocity fluctuations due to a lift-up process triggered by 
streamwise vortices. In fact, at the beginning of the energy growth
the instantaneous vorticity field shown in figure~\ref{fig:iso_D_growth}$(a)$
is characterized by the presence of quasi-streamwise vortices.
Successively the vortices give rise to instantaneous 
velocity profiles characterized by the typical ramp and cliff pattern, 
see e.g. \cite{pumir1994numerical,pumir1995persistent}.
The cliffs are regions of rapid increase of the streamwise velocity in the direction 
of the mean gradient in correspondence to spanwise vortex sheets, 
see figure~\ref{fig:iso_D_growth}$(b)$. These cliffs are associated to inflectional points of the velocity profiles or regions of high vorticity using the interpretation by Ellingsen and Palm, and therefore locally seeds of secondary instabilities.
Indeed the sheets become unstable and eventually roll up into 
spanwise vortices. 
The interaction of the spanwise vortices with the mean flow
generates the quasi-streamwise hairpin structures which induce
the observed energy bursts and the associated  large negative Reynolds 
stress. A similar scenario is presented in the simulations of \cite{BrandtLange2008} starting with the interactions of nearby laminar streaks.

\begin{figure}
\centering
\includegraphics[scale=0.33]{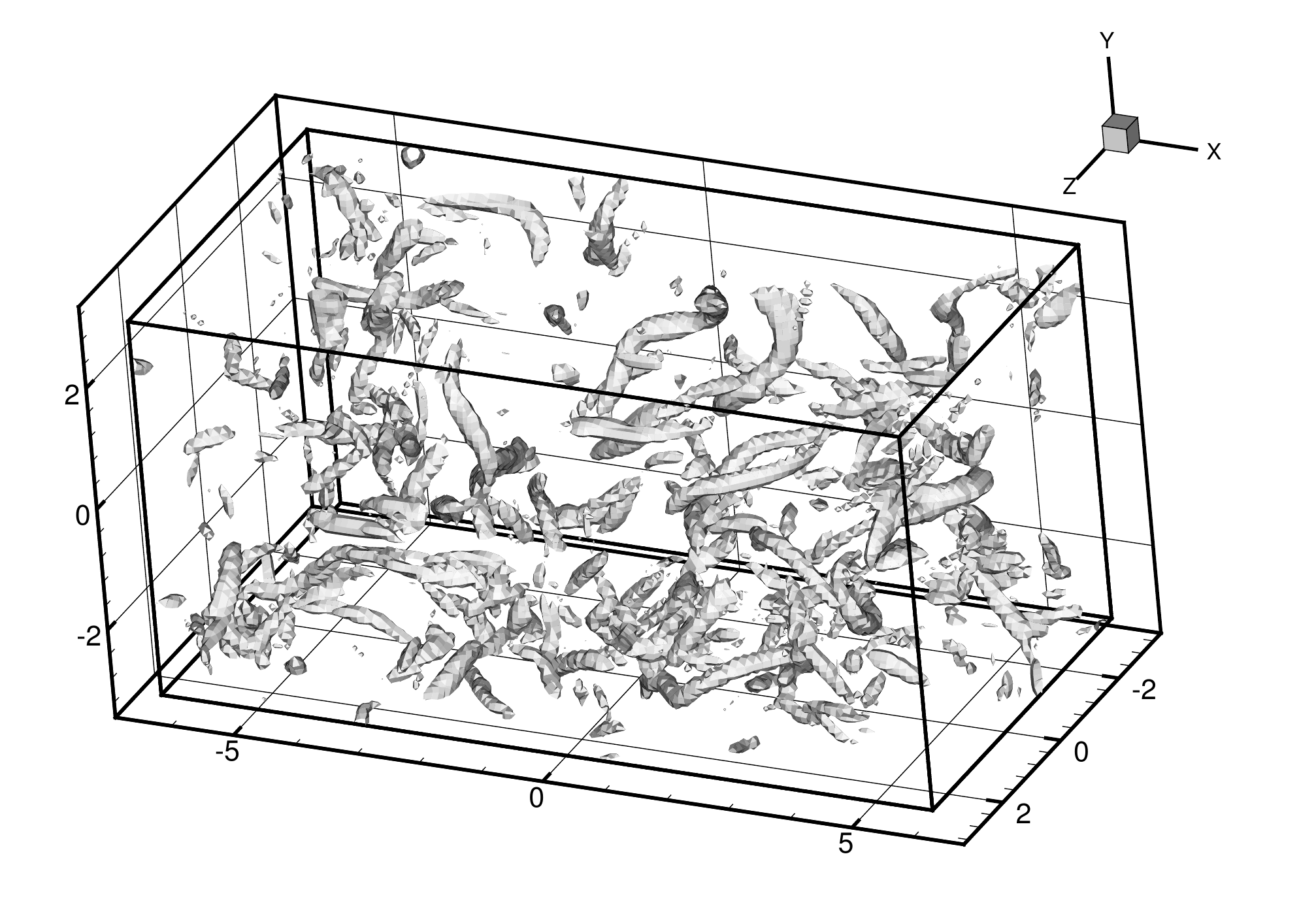}
\includegraphics[scale=0.33]{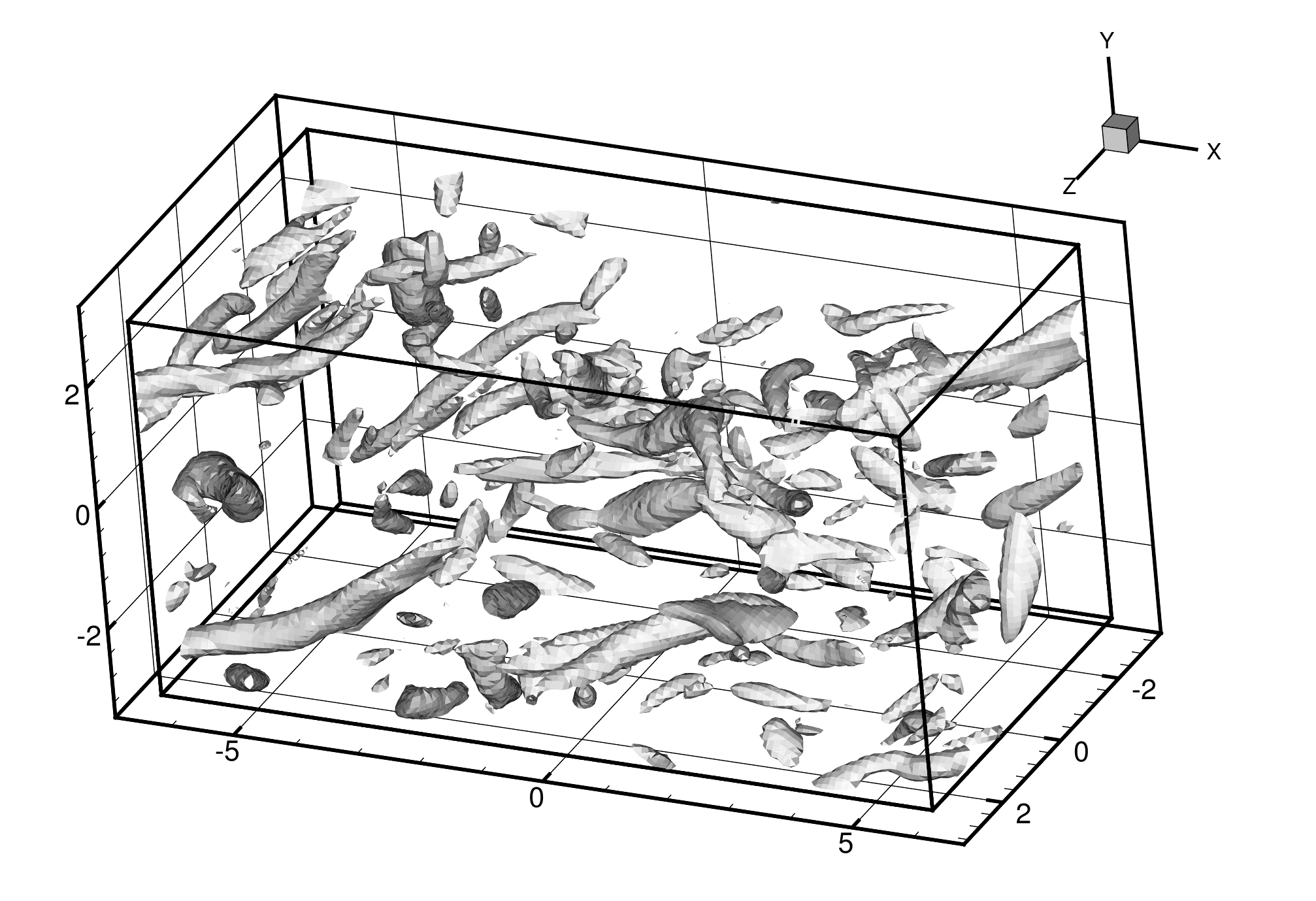}
\put(-390,100){{\large $(a)$}}
\put(-199,100){{\large $(b)$}}
\caption{$(a)$ Vortical structures after the energy maximum and $(b)$ in correspondence 
of the energy minimum during a cycle in turbulent homogeneous shear flow \citep[see also][]{gualtieri2002scaling}. 
\label{fig:iso_D_burst} }
\end{figure}

After the turbulent kinetic energy maximum, the non-linear interactions 
are overwhelming and the original ordered vortex structures turns into a randomized  
vorticity field, see figure~\ref{fig:iso_D_burst}$(a)$. Finally,
the mean flow align again the different flow structures,
figure~\ref{fig:iso_D_burst}(b), in correspondence of the energy minimum, 
just before the successive cycle starts. This is the process discussed by \cite{moffat67} and  \cite{phillips69}  using the linearized flow equations.
This dynamics, besides its relevance for 
turbulence understanding, strongly impacts the transport properties of turbulent 
shear flows. For instance, in particle laden flows, the coherent vortices
induce an anisotropic structure of the particle concentration 
field \citep{gualtieri2009anisotropic}, which in turn affects both 
inter-particles collisions \citep{gualtieri2012statistics} and turbulence 
modulation \citep{gualtieri2013modulation}.

The increase of the turbulent kinetic energy can be 
more quantitatively explained in terms of the linear lift-up mechanism and the related 
transient growth. A non linear mechanism is required instead to justify the saturation 
while the break-down of the ordered system of vortices might be explained by a secondary instability analysis. In Fourier space, 
as already discussed by \cite{pumir1996turbulence,gualtieri2002scaling}, the mode 
$(k_x,k_y,k_z)=(0,0,\pm 1)$ gives the leading contribution to the energy growth. 
The linearized evolution equation for this mode reads  
\begin{equation}
\label{linear_ns}
\left\{
\begin{array}{l}
\displaystyle  \frac{d \hat{u}}{dt}=-S \hat{v} - \nu \left(\frac{\pi}{\lambda_z}\right)^2\hat{u}\\ \\
\displaystyle \frac{d \hat{v}}{dt}= - \nu \left(\frac{\pi}{\lambda_z}\right)^2\hat{v} ,
\end{array}
\right. 
\end{equation}
where $S$ is the mean flow shear and $\lambda_z$ the spanwise width of the computational domain.
A growing amplitude is therefore expected whenever 
$S\left(Re(\hat{u}) Re(\hat{v}) + Im(\hat{u}) Im(\hat{v})\right)  < 0$. 
As the energy grows appreciably, the non-linear interaction with the 
other modes originate a strong energy transfer towards smaller scales. 
The characteristic frequency of the energy fluctuations is determined by the 
dynamic balance between growth of the basic mode and energy transfer. 
The saturation time, i.e. the time of the effective activation of the nonlinear 
energy transfer is related to the inviscid time scale $S^{-1}$, 
see e.g \cite{gualtieri2002scaling,yakhot2003simple}.
As a consequence, the bursting frequency roughly corresponds 
to the typical time of production of large scale velocity/vorticity instability which 
is the primary forcing mechanism of the homogeneous shear flow. 

\lb{As shown by the example above of the homogeneous shear flow,  the presence of a wall is therefore not necessary for the lift-up effect to sustain turbulence. In shear flows without any constraining walls but with inflectional velocity profiles, like free jets and plumes, however, inflectional instability also become important for turbulence generation \citep{Huerre98}.}

\section{The lift-up effect in non-Newtonian and multiphase flows}

In the first part of this article we have shown how the lift-up effect, or the algebraic growth of streamwise elongated modes, is a robust mechanism for disturbance amplification in shear flows and therefore a fundamental ingredient in the dynamics of transition and turbulence. Now, we would like to present recent as well as new results on the non-modal growth in complex fluids, i.e. fluids with a micro-structure that cannot be described by a linear relation between strain and stress.

\subsection{Inelastic fluids and viscosity stratification}

The effect of viscosity stratification and shear-dependent viscosity on instabilities and transition turbulence has been very recently reviewed by \cite{rama_annu}. Therefore, we briefly summarize some aspects related to the algebraic growth in parallel shear flows here, and refer the reader to this recent and comprehensive review.
As shear flows occur typically at high Reynolds numbers and are therefore expected to be
dominated by inertial effects, one would not expect viscosity variations to alter the instabilities significantly.
Viscosity contribute dissipation, but it is capable of altering the phase between the velocity fluctuations, the Reynolds stress, and thus the production of the disturbance kinetic energy. In addition, variations of the base flow, and in particular of the shear close to a wall, may have an impact on the lift-up effect.

The stability of inelastic non-Newtonian fluids has been studied extensively, and there is
consensus that shear-thinning is stabilizing and shear-thickening is destabilizing. The high-shear
regions are typically close to the wall, and shear thinning and thickening would make the velocity
profiles fuller and closer to being inflectional, respectively.  However, the effect on the transient growth is important in the linearly stable range of parameters and 
in flows where transition to turbulence follows the algebraic-growth route. In plane
Poiseuille flows of shear-thinning fluids, with viscosity perturbations ignored, transient growth is
slightly decreased \citep{ramaPRL}. With viscosity perturbations obtained approximately by
a simple shear-thinning model, transient growth is slightly increased \citep{nouarbottaro}, whereas
in Couette flow, transient growth is increased substantially in a shear-thinning fluid flow \citep{liu2011non}. Thus shear thinning, which damps the leading unstable mode, can promote turbulence
by non-modal mechanisms, still related to the lift-up.


\subsection{Viscoelastic fluids}

We now consider the case of viscoelastic fluids, fluids with a memory of past deformations. Among those, we will focus on dilute polymer suspensions that have been extensively studied due to the reduced drag in the turbulent regime \citep{White08}.

\begin{figure}
\begin{center}
\includegraphics[width=0.47 \textwidth]{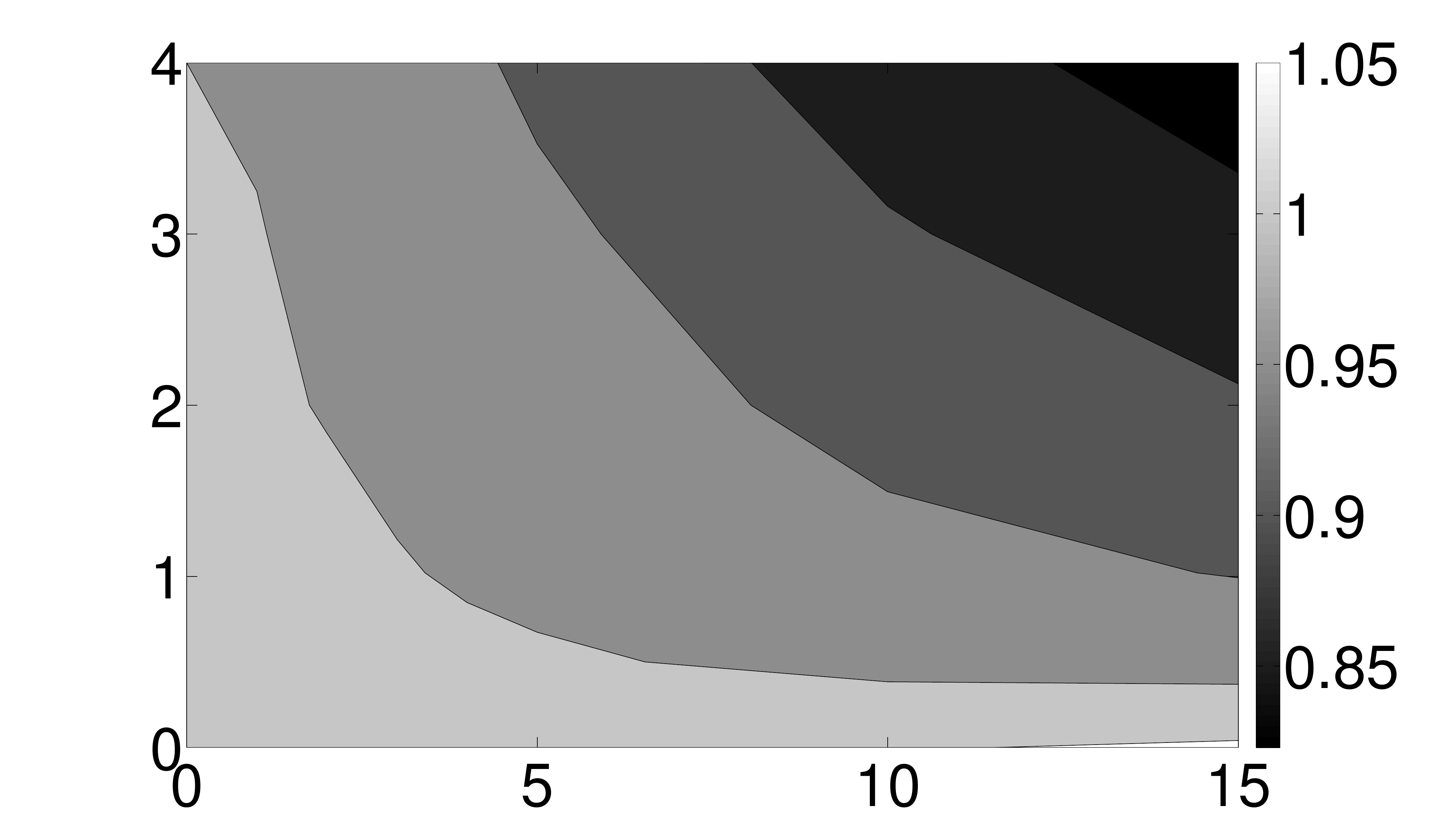} 
\includegraphics[width=0.47 \textwidth]{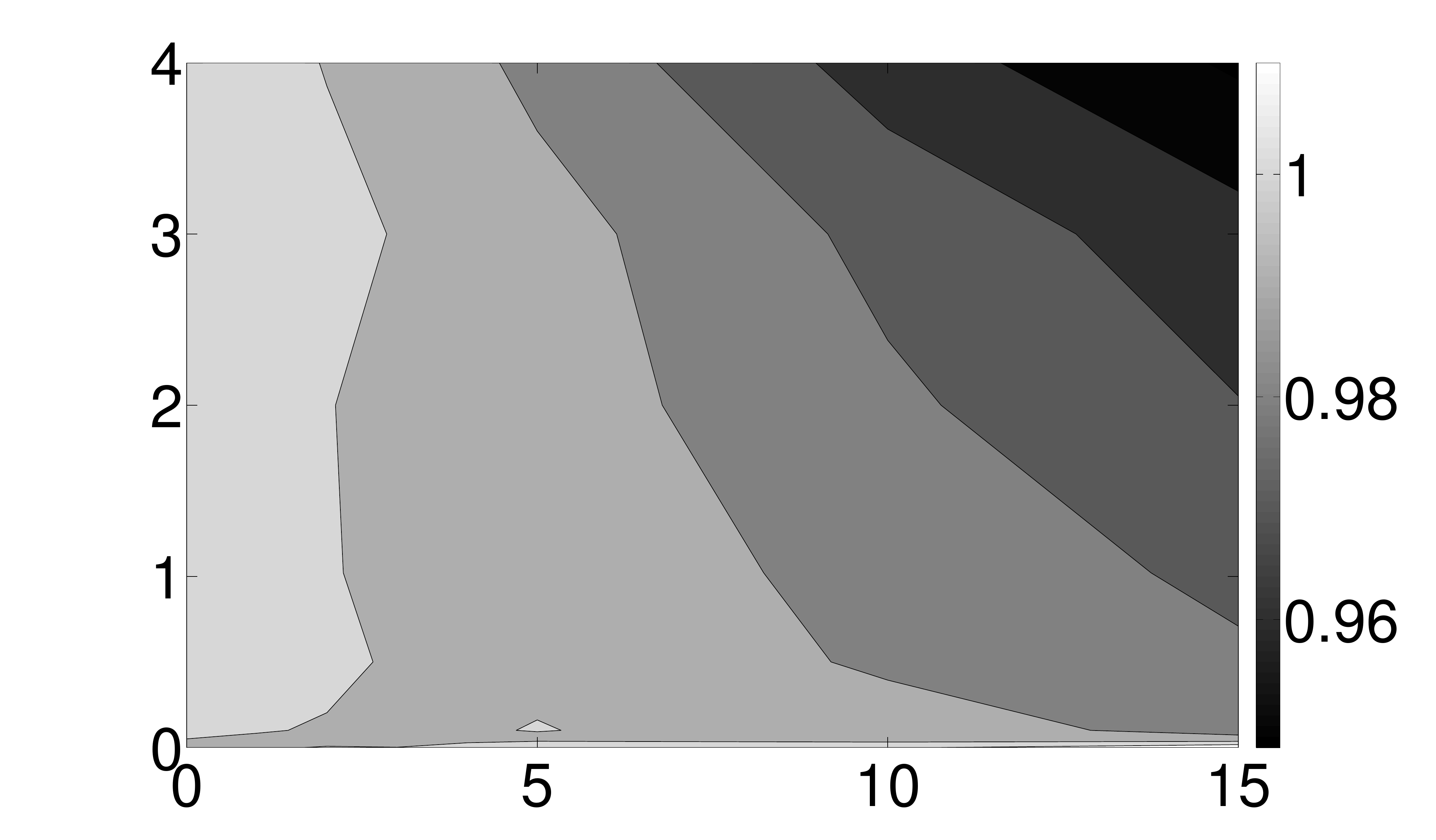}
\put(-380,90){{\large $(a)$}}
\put(-375,55){{$\alpha$}}
\put(-185,90){{\large $(b)$}}
\put(-180,55){{$\alpha$}}
\put(-290,-5){{$Wi$}}
\put(-100,-5){{$Wi$}}
\end{center}
\caption{$(a)$ Optimal transient growth $\max_t G(t)$ versus the streamwise wavenumber $\alpha$ and the Weissenberg number $Wi$, the non-dimensional polymer relaxation time,  normalized by the corresponding value for a Newtonian fluid. $(b)$ Time of maximum growth normalized by the corresponding value for Newtonian fluid. Poiseuille flow of a polymer suspension with Reynolds number $\Rey=4000$, viscosity ratio $\eta_s=0.9$ and polymer maximum extensibility $L=60$. All perturbations are characterized by a spanwise wavenumber $\beta=2$.}
\label{fig:fene-map}
\end{figure}
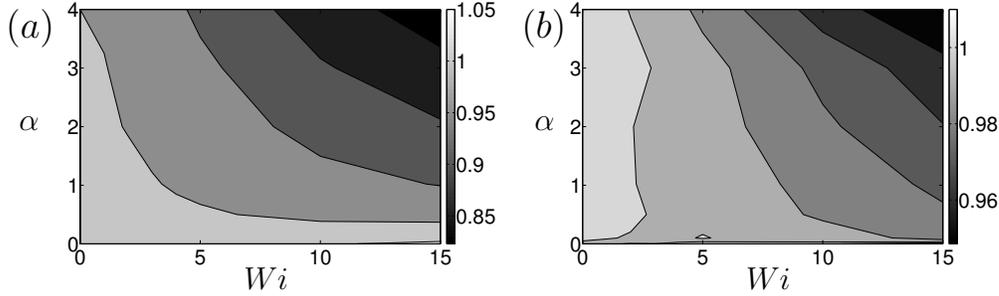

Receptivity analysis of channel flow of Oldroyd-B fluids at subcritical conditions is presented in \cite{Hoda08,Hoda09}, the latter work focusing on streamwise constant perturbations. There authors show how the lift-up effect, and the associated amplification of streaks, is still the dominant instability mechanism also in viscoelastic fluids. Here, we will discuss the non-modal analysis of Poiseuille flow of a polymer suspension modeled by the FENE-P closure \citep{bird1987}. A more detailed analysis is presented in \cite{mengqi}, where the effect of the polymer additives on the modal stability is also considered.

With the FENE-P model, the polymeric stress can be written as 
\begin{align}\bar{\tau}_p = \frac{f\bar{\textbf{C}}-I}{Wi},
\label{eq:fene}\end{align} 
where $Wi$ is the Weissenberg number defined as the ratio between the polymer relaxation time and the flow convective time scale and $f\equiv\frac{1}{1-\frac{\bar{C}_{kk}}{L^2}}$ is the Peterlin function, limiting the maximum polymer extensibility to $L$, with $\bar{C}_{kk}=\bar{C}$ the trace of the polymer conformation tensor. 
The non-dimensional constitutive equations for the evolution of the conformation tensor reads \citep[see e.g.][]{bird1987}
\begin{align}
\frac{\partial \bar{\textbf{C}}}{\partial t} + \bar{\textbf{u}}\cdot\nabla\bar{\textbf{C}} - \bar{\textbf{C}}\cdot(\nabla\bar{\textbf{u}}) - (\nabla\bar{\textbf{u}})^T\cdot\bar{\textbf{C}} = -\bar{\textbf{$\tau$}}_p,
\label{eq:oldcon}
\end{align} 
where $\bar{\textbf{$\tau$}}_p$ is related to the conformation tensor by equation (\ref{eq:fene}). 
The polymeric stress acts on the momentum balance as
\begin{equation}\label{eq:uN-S}
\frac{\partial{\bar{\textbf{u}}}}{\partial{t}} + (\bar{\textbf{u}}\cdot\nabla)\bar{\textbf{u}} = - \nabla \bar{p} + \frac{\eta_s}{Re}  \nabla ^2\bar{\textbf{u}} + \frac{1-\eta_s}{Re}\nabla\cdot\bar{\textbf{$\tau$}}_p,
\end{equation}
with the viscosity ratio $\eta_s$, the ratio between the solvent viscosity and the total viscosity.

Results of the linear behavior of plane channel flow of a polymer suspensions are reported in figure~\ref{fig:fene-map} for $\Rey=4000$, $\eta_s=0.9$, $L=60$ and fixed spanwise wavenumber $\beta=2$. In the panel $(a)$, we display the optimal transient growth versus the streamwise wavenumber $\alpha$ and the Weissenberg number $Wi$, normalized by the corresponding value for a Newtonian fluid.
The data indicate that the transient growth of streamwise independent modes, $\alpha \to 0$, is not affected by the polymer additives, even increased at the largest $Wi$ considered. Interestingly, the amplification of streamwise dependent modes, finite $\alpha$'s, is reduced in the presence of polymers, up to about 20\%.  Note once more that the largest absolute values of energy growth are observed for low values of $\alpha$, the figure reporting solely the variations due to the polymers.
The time of maximum growth, normalized by the corresponding value for Newtonian fluid, is shown in figure~\ref{fig:fene-map}$(b)$: very weak modifications are observed in this case. In summary, the lift-up effect is still the responsible for the largest energy growth; recalling that the streak growth occurs on a long time scale, we find a weak destabilization when the time scale of the instability is longer than the polymer relaxation time, while the flow is more stable when $Wi$ is of the order of the time over which the amplification is observed (finite values of $\alpha$). 

\begin{figure}
\begin{center}
\includegraphics[width=0.55 \textwidth]{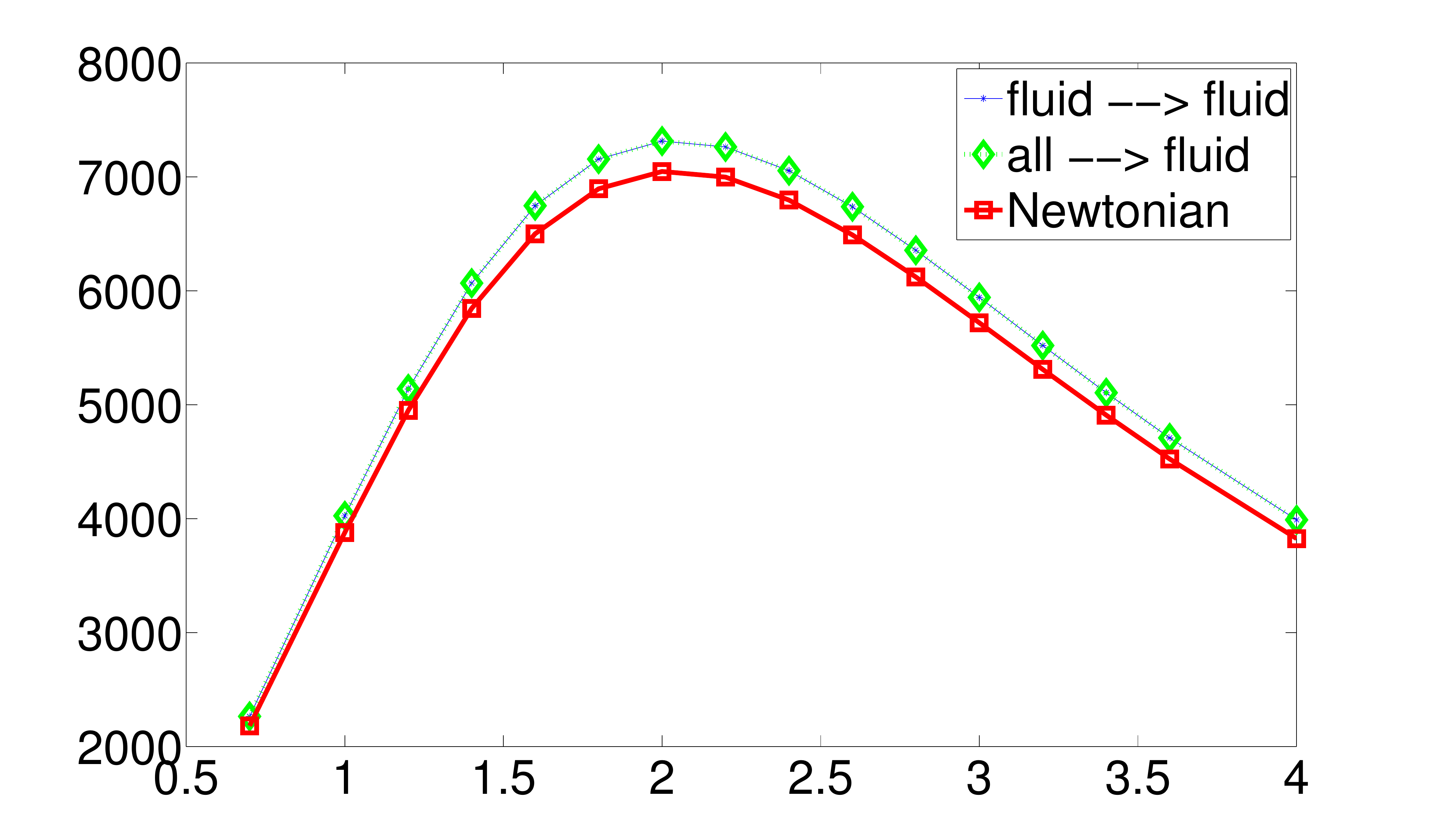} 
\put(-240,75){{$G(t)$}}
\put(-110,-5){{$\beta$}}
\end{center}
\caption{Optimal transient growth $\max_t G(t)$ versus the spanwise wavenumber $\beta$ for Poiseuille flow of a polymer suspension with $\Rey=6000$, Weissenberg number $Wi=0.9$, viscosity ratio $\eta_s=0.9$ and maximum extensibility $L=60$. All perturbations are characterized by a streamwise wavenumber $\alpha=0$. See text for the legend explanation. }
\label{fig:fene-ic}
\end{figure}

We would like to conclude by mentioning that the formation of streak is also found in inertialess polymer suspensions. Transient growth analysis of inertialess Couette and Poiseuille flow of viscoelastic Oldroyd-B fluids is presented in \cite{Jovanovic2010,Jovanovic2011}. 
The wall-normal fluctuation of the polymer stress generates the largest transient growth and the stream-wise component is the most sensitive to the elasticity in the case of weak inertia. The stretching of the polymer molecules results in a lift up of the disturbances similar to that observed in inertia-dominated Newtonian flows. The effect of this type of initial perturbation on the transient growth of the streaks in the inertial regime is reported in figure~\ref{fig:fene-ic}. In the legend, the notation $all \to fluid$  indicates that the perturbation at time $t=0$ is acting both on the velocity field and polymer conformation while the output energy is measured by the fluid kinetic energy only, and similarly for the other curves displayed.
The energy growth is larger than in the Newtonian fluid and of the same magnitude whether the polymer are initially stretched or not.
The case $polymer \to fluid$, not shown in the figure, gives a negligible growth at the relatively high Reynolds numbers considered here, $\Rey \approx \mathcal{O}(10^3)$; in other words, polymer stretching is not able to trigger disturbance growth when fluid inertia is relevant.

\subsection{Channel flow with small particles}

We examine next the non-modal linear stability of the plane channel flow of dilute particle suspensions where particles are assumed to be solid, spherical, heavier than the fluid phase and smaller than the flow length scales;  under these assumptions, the coupling between particle and fluid flow is modeled by the Stokes drag only. The lift-up process in a channel laden with finite-size neutrally buoyant particles is presented in the next section.
The dynamics of small inertial particles transported in a flow is crucial in
many engineering and environmental applications.   
Interestingly,  adding dust to a fluid may reduce the drag in pipe flows \citep{Sproull1961}. To explain this phenomenon it has been suggested that the dust delays transition and dampens the formation of turbulent structures. 

A detailed investigation of the modal instability of dusty gases is presented in \cite{klinkenberg2011modal}. As shown in earlier studies \citep{Saffman62,Michael64},
exponentially growing perturbations show the potential for a significant stabilization, i.e.\ an increase of the critical Reynolds number. The largest stabilization is observed when the ratio between the particle relaxation time and the frequency of the wave is of order one. After  examining the energy budget this stabilization is attributed to the increase of the dissipation in the flow caused by the Stokes drag.

\begin{figure}
\begin{center}
\subfigure {\includegraphics[width=0.44\textwidth]{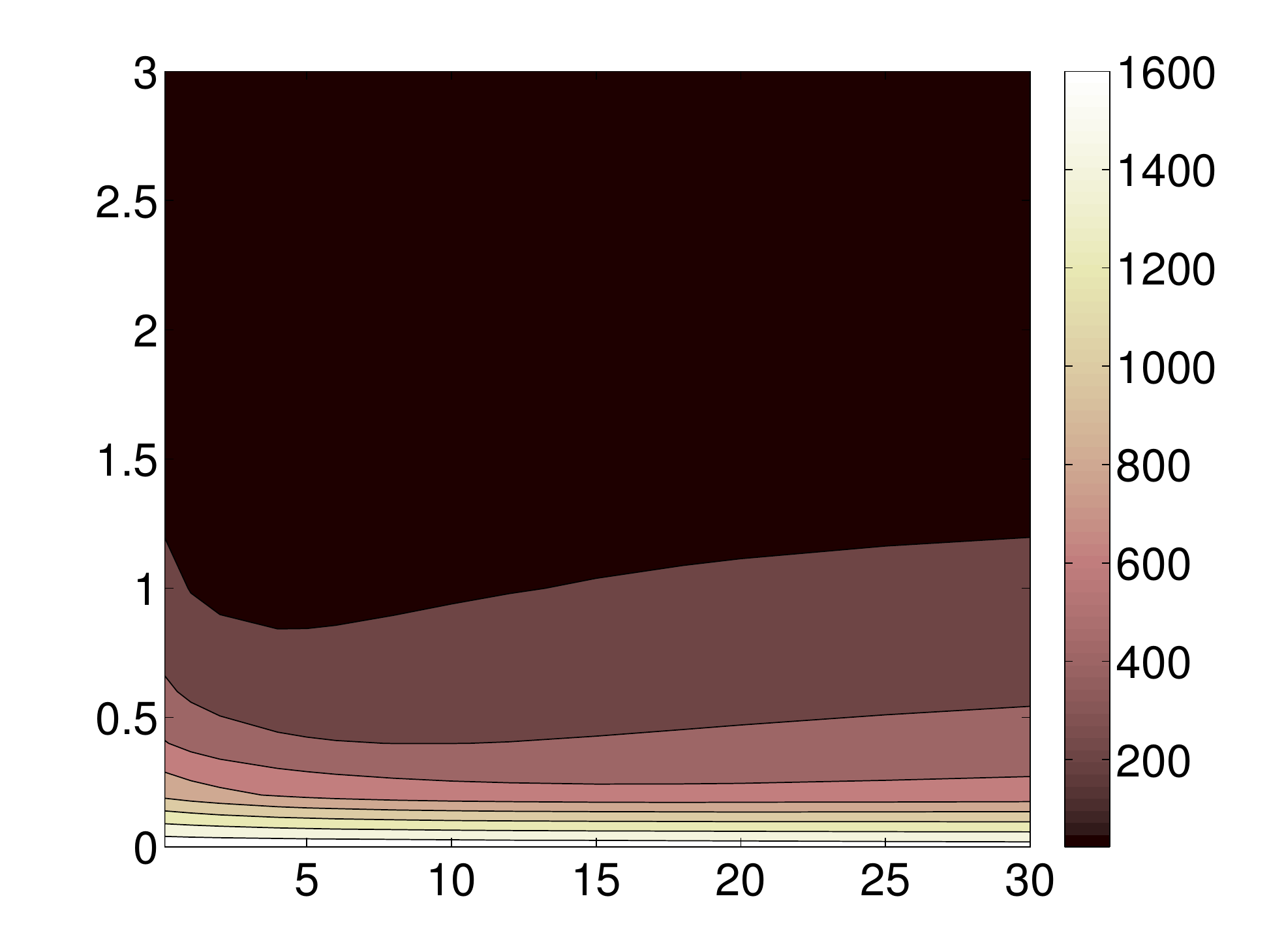}}
\put(-180,110){{\large $(a)$}}
\put(-175,65){{$\alpha$}}
\put(-100,-5){{$SR$}}
\subfigure {\includegraphics[width=0.44\textwidth]{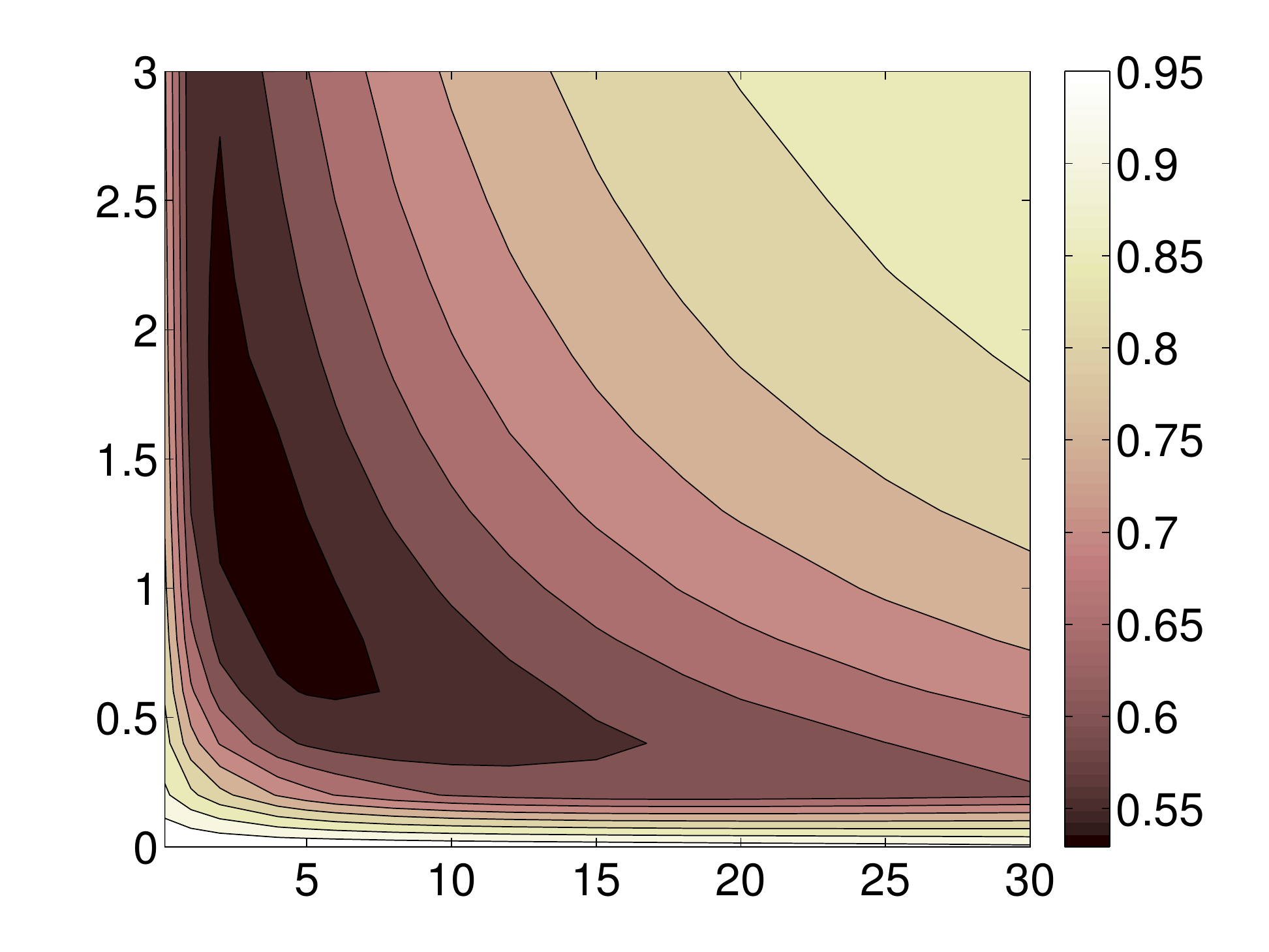}} 
\put(-180,110){{\large $(b)$}}
\put(-175,65){{$\alpha$}}
\put(-100,-5){{$SR$}}

\subfigure {\includegraphics[width=0.44\textwidth]{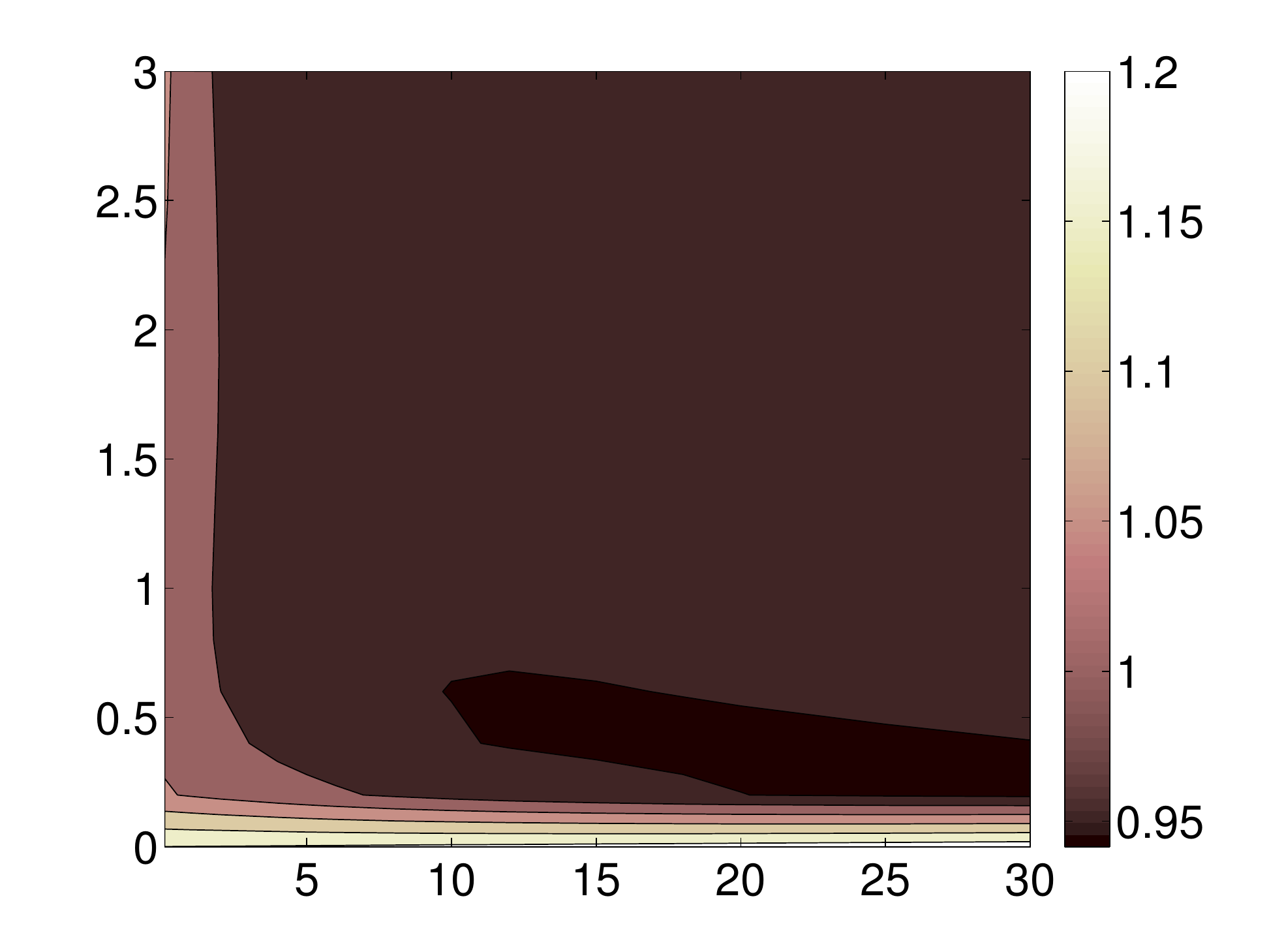}} 
\put(-180,110){{\large $(c)$}}
\put(-175,65){{$\alpha$}}
\put(-100,-5){{$SR$}}
\end{center}
\caption{$(a)$ Optimal transient growth $\max_t G(t)$ versus the streamwise wavenumber $\alpha$ and the Stokes number $SR$ for particle-laden channel flow at $\Rey=3000$. $(b)$ Same as in $(a)$ normalized by the corresponding value for a Newtonian fluid. $(c)$ Time of maximum growth normalized by the corresponding value for Newtonian fluid. All perturbations are characterized by a spanwise wavenumber $\beta=2$. The mass fraction is $f=0.2$.}
\label{fig:part-map}
\end{figure}
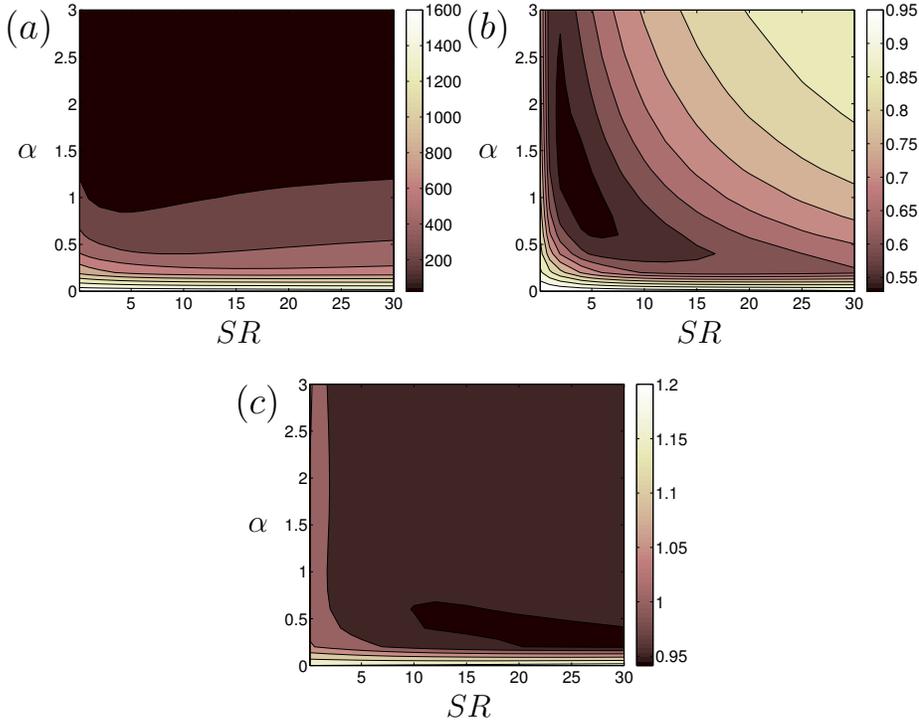

In our analysis, we adopt the continuous, or Eulerian, model as \cite{Saffman62} \lb{and corresponding nomenclature}: the particles are assumed to be under the
action of the Stokes drag only, lift force, buoyancy and added mass are neglected. Denoting $K=6\pi a \mu$ the Stokes drag \lb{per relative velocity}, with $a$ the particle radius and $\mu$ the fluid viscosity, the particle relaxation to the flow velocity occurs on a time scale $s_p=m_p/K$ where $m_p$ is the mass of a single particle. In the limit of high density ratios, particle density over fluid density, the non-dimensional parameters defining our problem are the Reynolds number, based on the fluid velocity and the half-channel width $\Rey= \rho U h/ \mu$, and, in addition, the mass fraction $f=m_p/m_f$,  and the particle relaxation time $S_p=\nu s_p/ h^2$. The non-dimensional particle relaxation time based on the flow convective time scale is $SR= S_p \cdot \Rey$.

The linearized stability equation for particle-laden channel flow, casted as an initial value problem, can be found in  \cite{klinkenberg2011modal}. Note that for channel flow, the base flow is the parabolic Poiseuille profile also in the presence of the solid phase. 
This earlier study focuses mainly on the amplification of streamwise independent disturbances which  are shown to still be the most amplified. Here, we consider the behavior of streamwise dependent and independent modes of fixed wavenumber $\beta=2$, as in the previous section for a dilute polymer suspension.

The overall maximum energy growth, $\max_t G(t)$, is reported in figure~\ref{fig:part-map}$(a)$ versus the streamwise wavenumber $\alpha$ and the non-dimensional relaxation time $SR$ for $\Rey=3000$, $\beta=2$ and $f=0.2$. In the limit  $SR \to 0$ we find passive tracers, while inertial effects become more and more important at higher values of $SR$ (the ballistic limit).  As in Newtonian fluids, the most dangerous perturbations are those with $\alpha \to 0$. The modifications induced by the solid phase on the disturbance amplification are displayed in figure~\ref{fig:part-map}$(b)$ where the data in $(a)$ are divided with the corresponding maximum energy growth in the flow without particles.
The streak transient growth is basically unaffected by the presence of the solid phase, while a significant stabilization is observed for $2<SR<8$. This attenuation is found to occur for values of the particle relaxation times comparable with the time over which the transient disturbance growths observed. The same physical mechanisms identified in \cite{klinkenberg2011modal} for model instabilities, increased dissipation, seems to be at work also for non-modal instabilities. The relative variation of the time at which the maximum growth is attained is displayed in figure~\ref{fig:part-map}$(c)$: the transient growth is delayed proportionally to the particle volume fraction while the growth of oblique modes is slightly shorter.

\begin{figure}
\begin{center}
\includegraphics[width=0.55\textwidth]{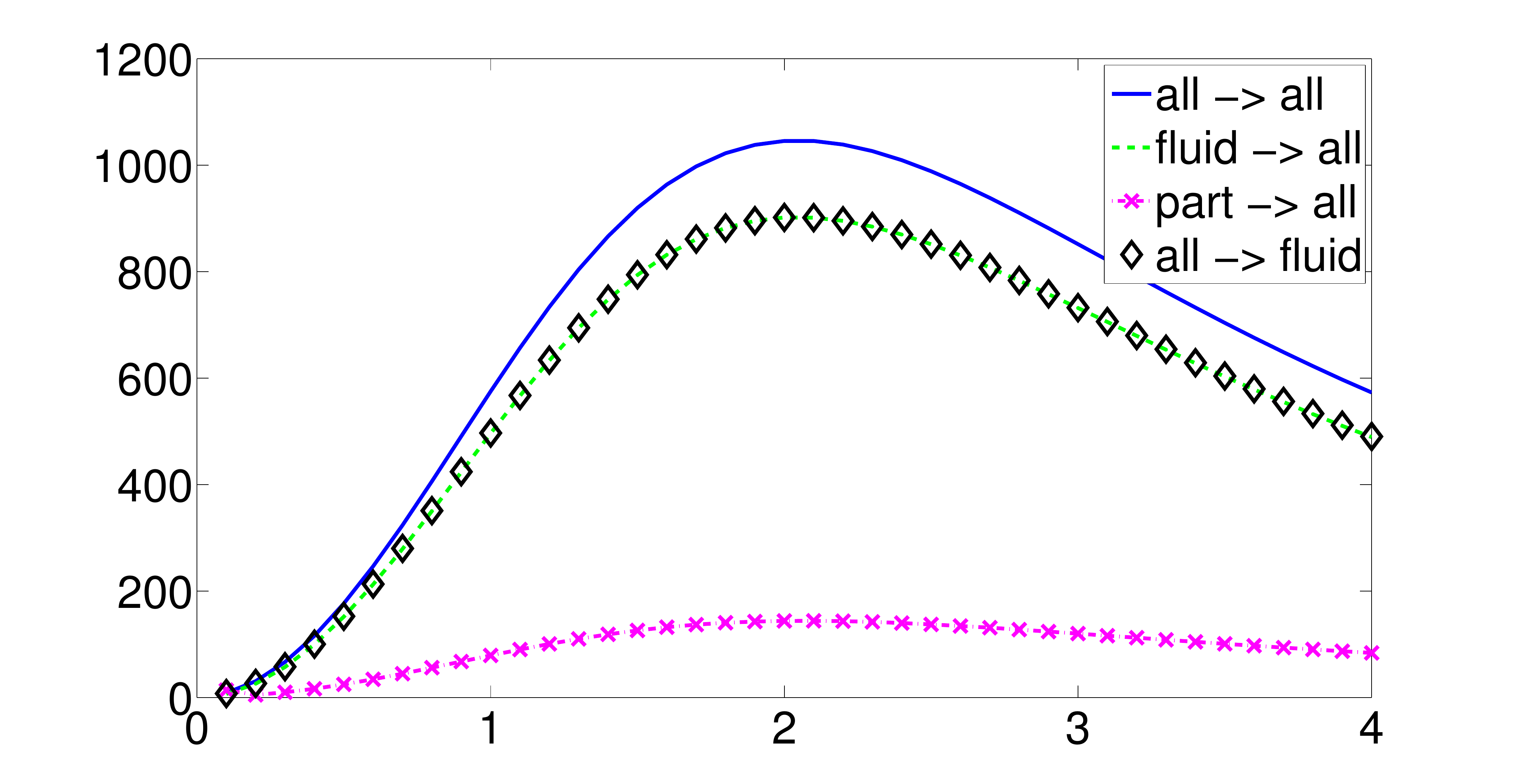}
\put(-238,75){{$G(t)$}}
\put(-110,-8){{$\beta$}}
\end{center}
\caption{Optimal transient growth $\max_t G(t)$ versus the spanwise wavenumber $\beta$ for a channel flow laden with small particles. \lb{The flow Reynolds number  $\Rey = 2000$, the viscous particle relaxation time $S_p = 0.0025$ and mass fraction $f = 0.15$}. The optimal initial condition consists of fluid or particle velocity or both, denoted respectively as ÓfluidÓ, ÓpartÓ and ÓallÓ. The output is measured by the fluid disturbance kinetic energy or by the total kinetic energy of the mixture. All perturbations are characterized by a streamwise wavenumber $\alpha=0$.}
\label{fig:Gpart}
\end{figure}

The results in figure~\ref{fig:part-map} are obtained perturbing initially only the fluid velocity and measuring the fluid kinetic energy at later times. The flow behavior in the case of different initial perturbations is depicted in figure~\ref{fig:Gpart} where we focus on disturbances characterized by the streamwise wavenumber $\alpha=0$. If perturbing the fluid and particle velocity at time $t=0$, we find an increase of the energy growth with respect to the single fluid case by a factor $(1+f)^2$; a decrease by a factor $f$ is instead observed if only the particle velocity is initially perturbed.

The generation of streamwise streaks via the lift-up mechanism is therefore the dominant disturbance-growth mechanism in canonical shear flows at subcritical conditions also in the presence of small heavy particles: the length scales of the most dangerous disturbances are unaffected while the disturbance growth is slower. 
The increase of the non-modal amplification scales with the particle mass fraction, $(1+f)^2$, something which can be explained in terms of the
ratio between the particle relaxation time, $SR$, and the time of optimal growth, $t_{max} \approx \Rey$.
This ratio assumes very low values in the case of the low-frequency non-modal growth of the streaks, $SR/\Rey \approx S_p$ the relaxation time based on the viscous time scale,
and therefore the effect of particles is just that of altering the fluid density (from here the factor $f$). Particles have the time necessary to follow the slow formation of the streaks: the particles increase the solution density and the Reynolds number of the laden fluid becomes  $\Rey_s = (1 + f ) \Rey$. As the optimal growth in unladen flows is proportional to $\Rey^2$, the presence of the particles increases the energy gain by $(1+f)^2$.

The effect of the particle on the full transition process is discussed in \cite{klinkenberg2013numerical}, where two bypass scenarios, both characterized by the streak growth, are investigated. In this study, the authors show that the solid phase alters the transition to turbulence, and arguably also the self-sustaining turbulent cycle, 
in two ways. First, by reducing the amplification of the oblique modes particles  make the streak more persistent and delay their breakdown: the flow is less noisy and the streamwise streaks are even more dominant in the flow. Second, the particles reduce the amplitude of oblique modes and thus their nonlinear interactions necessary to generate new streaks. 
In summary, particles do affect the transition to turbulence and the turbulence \citep[see the drag reduction studies in e.g.][]{Zhao2010} not by altering the lift-up effect but rather by modifying the dynamics of the oblique waves necessary for the streaks regeneration and breakdown.

The modal and non-modal stability of small spherical particles of density of the order of that of the fluid is reported in \cite{klinkenberg2013light}. Here, the Stokes drag, added mass and fluid acceleration are used to describe the interactions with the fluid. The results indicate that the inclusion of the extra interaction
terms does not induce any large modifications of
the subcritical instabilities in wall-bounded shear flows. The analysis of the Basset history force  \citep{MaxeyRiley83}   shows that this force has a negligible effect on the flow instability. Excluded volume effects, finite-sized particles,  may however have an impact on the flow stability as discussed in the next section.

\subsection{Finite-size neutrally-buoyant particles}

\begin{figure}
\begin{center}
\includegraphics[trim = 20mm 61mm 30mm 60mm, clip, width=0.9\textwidth]{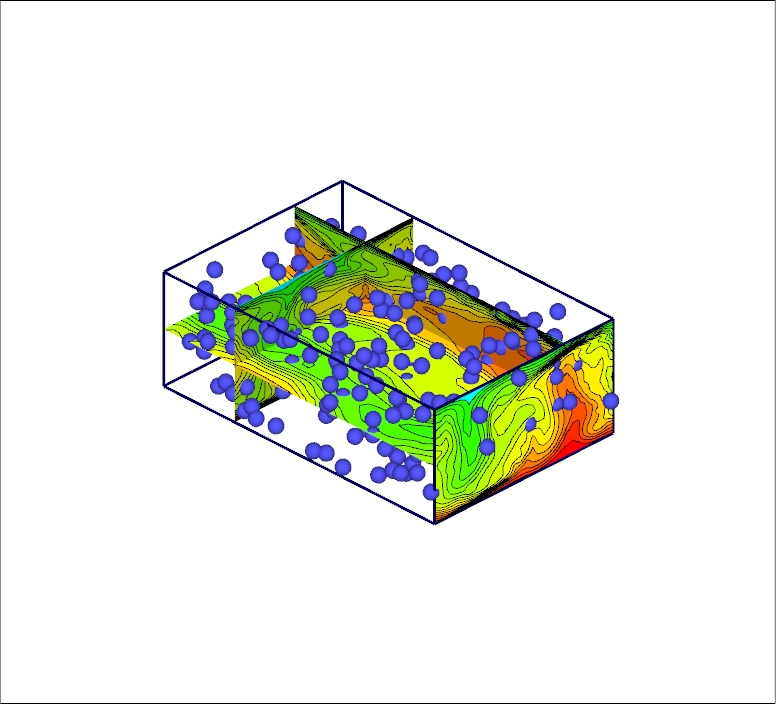}
\end{center}
\caption{ Snapshot of particle-laden Couette flow at Reynolds number $\Rey=600$ and volume fraction $\phi=0.05$ during the transition to turbulence. For the sake of clarity only 50\% of the particles are displayed.}
\label{fig:visua}
\end{figure}


We conclude our survey on the algebraic instability of non-Newtonian flows by examining the case of a suspension of neutrally buoyant finite-size particles, i.e.\ particles whose radius $a$ is comparable with the flow scales. In particular, we shall examine Couette flow at two Reynolds number, $\Rey=350$ and $\Rey=600$ in order to  first document the effect of the solid phase on the transient growth of the streaks and then the complete transition to turbulence.
The ratio between the channel half width $h$ and $a$ is $h/a=18$ and $h/a=10$, respectively. The Reynolds number is based on the velocity of the two walls, $U_{wall}$, and the half channel width. 

The simulations are performed with the Immersed Boundary Method to fully resolve the coupling between the fluid and solid phase; in this particular implementation a Lagrangian mesh is used to track the particle surface in the fluid. 
The numerical algorithm coupling finite-sized moving particles with the surrounding fluid flow 
was originally  developed by \cite{Uhlmann} and later expanded by \cite{Breugem2}, 
to model the interactions of multiple inert spheres with second order accuracy. The near field interactions and collisions between neighboring particles and
solid surfaces are modeled using both a corrective lubrication force and the soft-sphere
collision model. The method was recently used to investigate the rheology of semi-dilute and dense suspensions in the inertial regime \citep{picano_prl} as well as active suspensions \citep{lambert2013active}, where several validations have been performed. 

The optimal initial condition for the single phase fluid of wave vector $\alpha=0$ and $\beta= 2\pi/Lz$, with $L_z$ the spanwise dimension of the box,  is imposed at $t=0$.
For the case of higher $\Rey$, a three-dimensional disturbance of wave vector $\alpha=2 \pi/L_x$ and $\beta= 2\pi/Lz$ is also forced to trigger the transition to turbulence; its energy is 1/9 of that of the streaky mode \citep{duguet2010towards}.
Periodic boundary conditions are implemented in the streamwise and spanwise direction whereas Dirichlet conditions, $U=\pm 1$, are imposed at the walls.
As mentioned above two configurations are examined: i) $\Rey=350$ and domain size of $3 h \times 2h \times 3h $ in the streamwise, wall-normal and spanwise directions; and ii) $\Rey=600$ and domain size of $6 h \times 2h \times 4h $, the latter to have sustained turbulence. 
The resolution is $432 \times 288 \times 432$ grid points for the first and $480 \times 160 \times 320$ in the second configuration; in each case we used 16 points per particle diameter.
A visualization of the flow is presented in figure~\ref{fig:visua} for the second configurations and the highest volume fraction considered, $\phi=0.05$; only half of the particles are displayed for clarity.

\begin{figure}
\begin{center}
\includegraphics[width=0.45 \textwidth]{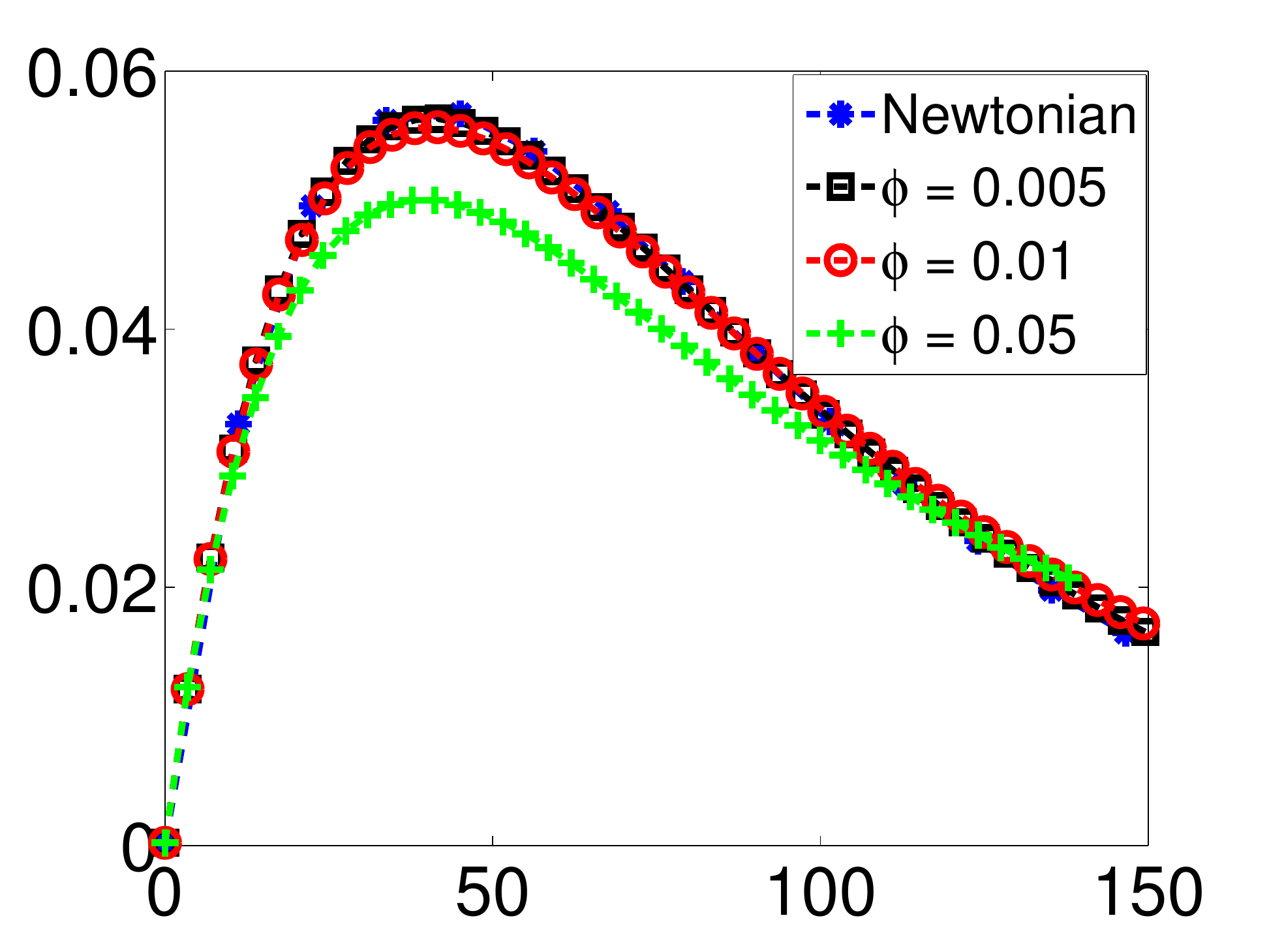} 
\includegraphics[width=0.45 \textwidth]{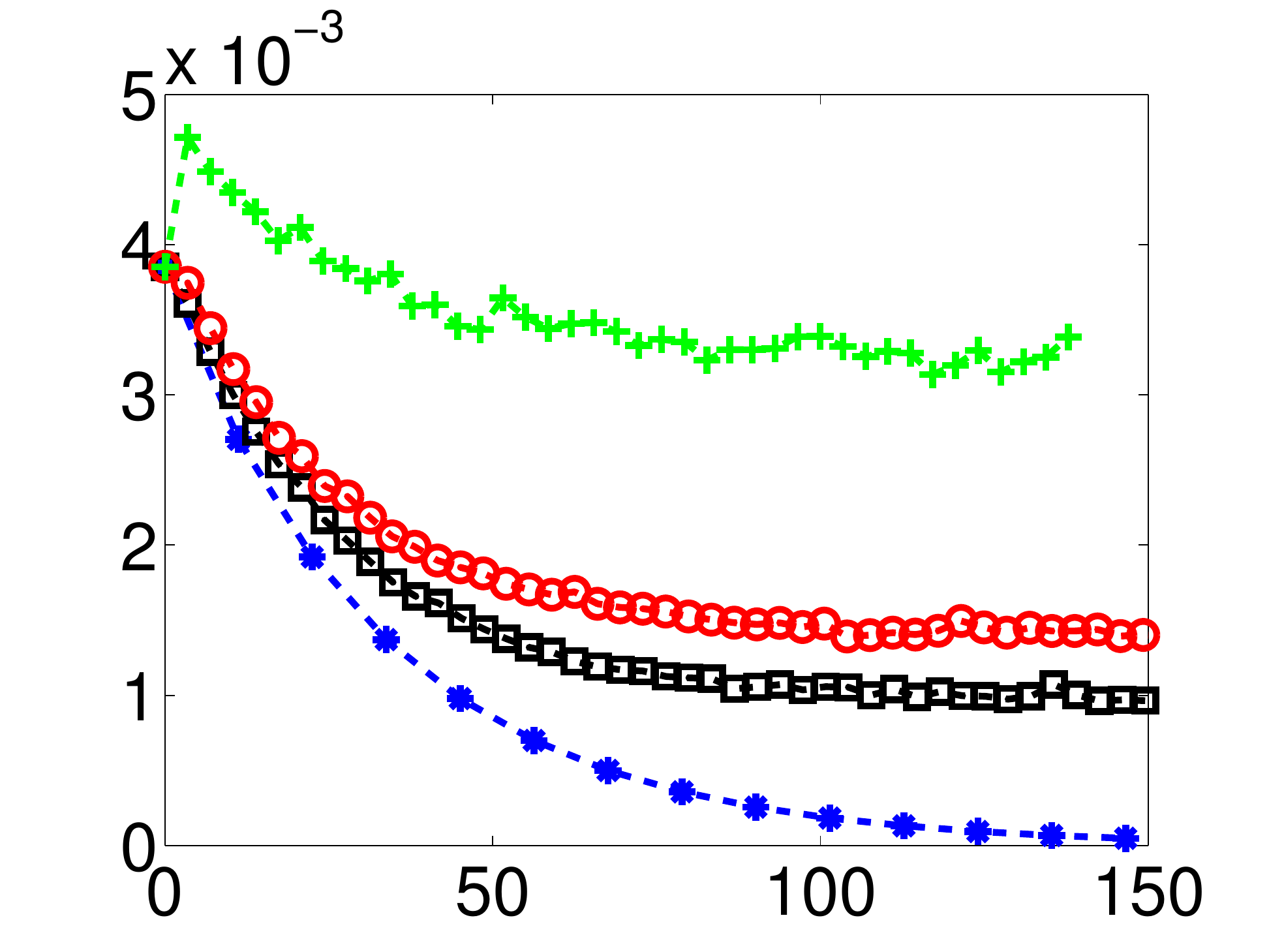} 
\put(-375,110){{$(a)$}}
\put(-370,70){{$u_{rms}$}}
\put(-275,0){{$t$}}
\put(-185,110){{$(b)$}}
\put(-185,70){{$v_{rms}$}}
\put(-85,0){{$t$}}
\end{center}
\caption{ Time evolution of the optimal initial disturbance in particle laden Couette flow at $\Rey=350$ with domain size $3 h \times 2h \times 3h$ and particle radius $a=h/18$. $(a)$ Streamwise velocity perturbation.
$(b)$ Wall-normal velocity perturbation for different values of the volume fraction $\phi$.}
\label{fig:rms_lowRe}
\end{figure}

The disturbance growth for the configuration with lower Reynolds number is displayed in figure~\ref{fig:rms_lowRe}. Here, the disturbance is measured as the integral over the computational domain of each of the velocities where the base Couette profile is subtracted from the streamwise component.
The initial condition is chosen to give a moderate streak amplitude at the time of maximum growth, $\Delta U \approx 0.27 U_{wall}$, so that nonlinear effect are not important, secondary instability far to come, and the disturbance evolution is close to the linear case in the absence of particles (denoted as Newtonian flow in the figure). 
The curves of the $u_{rms}$ clearly indicate that the transient growth of the streaks is still evident in the flow, with a reduction of about 15\% only for the highest volume fraction considered. The $v_{rms}$ decays monotonically in the Newtonian case as streamwise independent disturbances are only subject to viscous forces. The finite-size particles act as an extra localized forcing to the flow and the decay of cross-stream disturbances is therefore slower and almost completely quenched when increasing the particle volume fraction $\phi$. Flow visualizations around the time of maximum $u_{rms}$ show a pair of streamwise streaks of more irregular shape in the presence of the solid phase. Interestingly, the final decay of the streaks, $t>120$, is slower due to the continuous forcing by the particles, and the streamwise correlation of the structures decreases.

\begin{figure}
\begin{center}
\includegraphics[width=0.45 \textwidth]{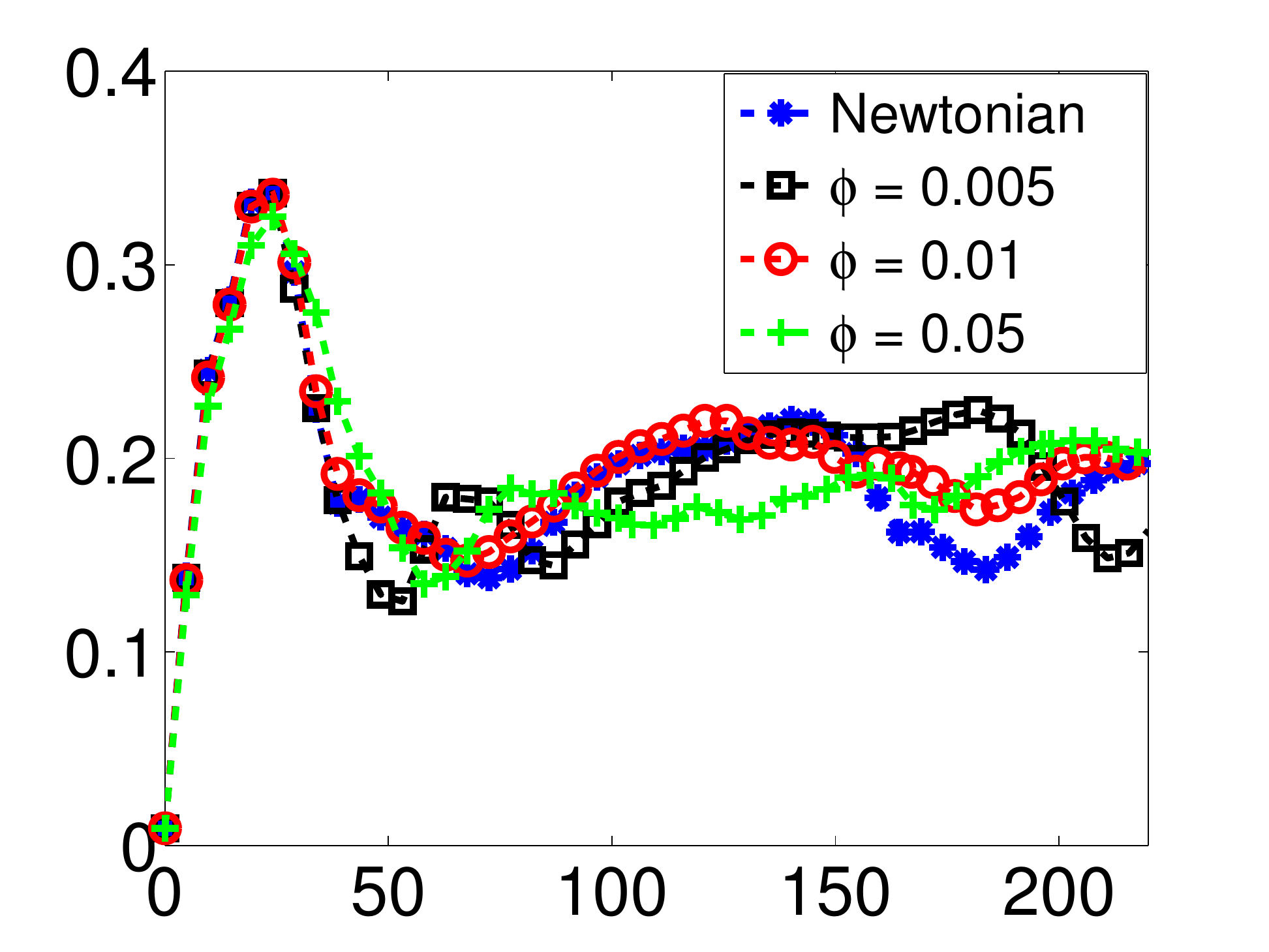} 
\includegraphics[width=0.45 \textwidth]{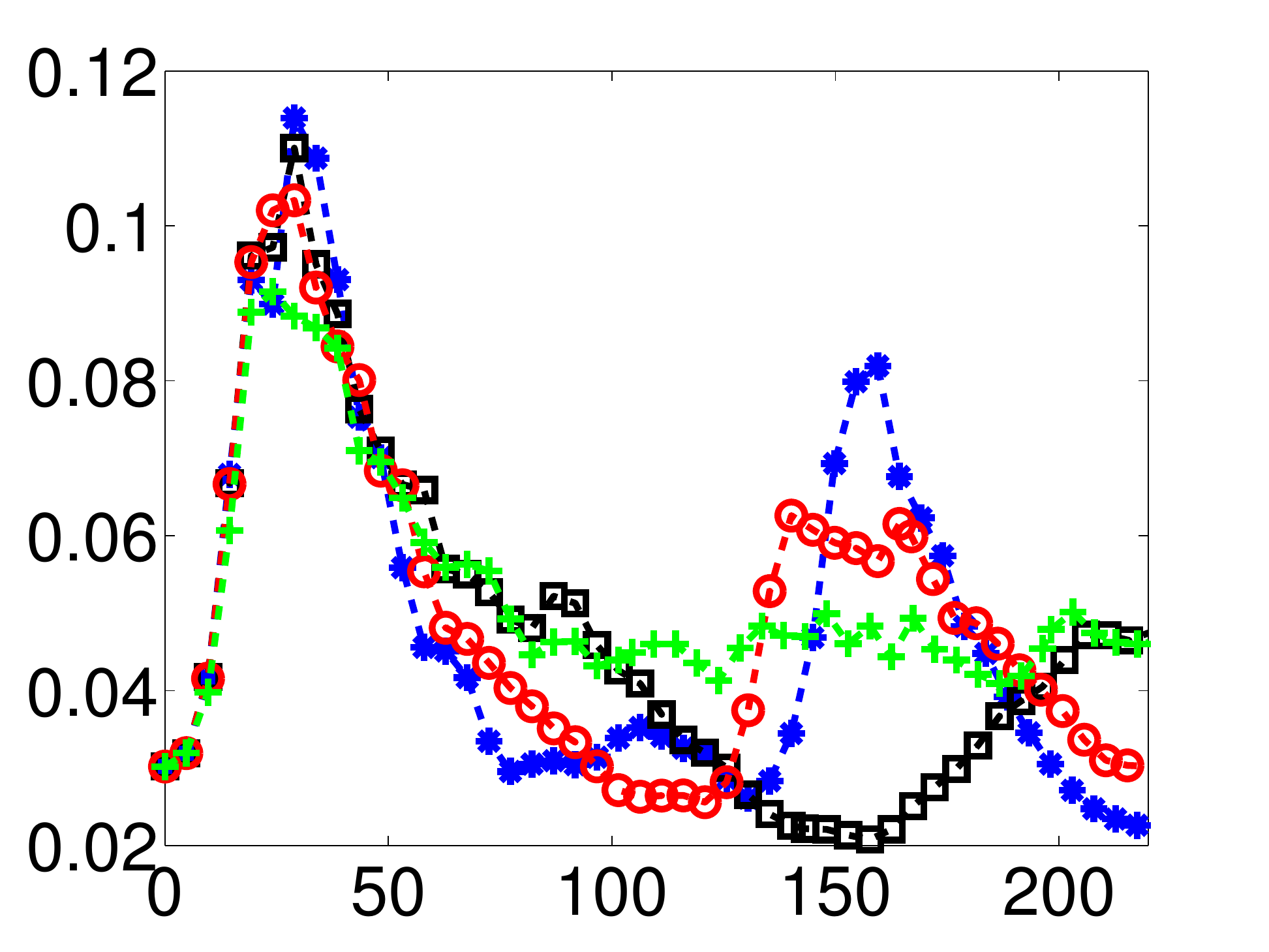} 
\put(-375,110){{$(a)$}}
\put(-375,70){{$u_{rms}$}}
\put(-275,-10){{$t$}}
\put(-188,110){{$(b)$}}
\put(-190,70){{$v_{rms}$}}
\put(-89,-10){{$t$}}
\end{center}
\caption{ Time evolution of the disturbance energy during transition to turbulence in particle laden Couette flow at $\Rey=600$ with domain size $6 h \times 2h \times 4h$ and particle radius $a=h/10$. $(a)$ Streamwise velocity perturbation.
$(b)$ Wall-normal velocity perturbation for different values of the volume fraction $\phi$.}
\label{fig:rms_highRe}
\end{figure}

Next, we examine the flow at higher Reynolds number and with larger particles, $h/a=10$, to see whether the presence of the particles, weakly affecting the lift-up process at moderate disturbance amplitudes, has an effect on the full transition process as done in \cite{klinkenberg2013numerical} for small particles. The results from these simulations are displayed in figure~\ref{fig:rms_highRe}.
The presence of the optimal oblique mode explains the initial growth of $v_{rms}$; the time of transition is not altered by the presence of a solid phase and the level of fluctuations in the turbulent regime is also weakly dependent on the volume fraction, at least for the cases considered here. The transition to turbulence follows therefore the bypass scenario: growth of streamwise streaks and secondary instability once they have reached high enough amplitude.
The flow at the time of maximum $u_{rms}$ for the case $\phi=0.05$ is displayed in figure~\ref{fig:visua} to show the bending of highly irregular streaks prior to breakdown \citep[cf.][]{cossu2011secondary}.

It is also interesting to note that the oscillations of $v_{rms}$, indicative of the different stages of the turbulence self-sustaining cycle, become less and less evident when increasing the volume fraction $\phi$. This is an indication of a change in the structure of the turbulent flow of a suspension and will be discussed elsewhere.  To conclude, we note that the presence of finite size particles does not significantly affect the lift-up process and therefore the following bypass transition to turbulence, confirming once more how this is a very robust process in shear flows, leaving a "permanent scar" in the flow \citep{Landahl75}. As next steps, it will be worthy investigating the flow behavior at higher volume fractions (a challenge from a computational point of view) to characterize how transition changes in the dense regime. The experiments by \cite{Matas2003PFL} in a particle-laden pipe flow show a non-monotonic behavior of the transitional Reynolds number when increasing $\phi$, a fact that cannot be explained solely by the increase of the system viscosity due to the presence of the particles.

\section{Conclusions}

Elongated flow structures consisting of positive and negative streamwise perturbation velocity alternating in the spanwise direction, the streaks, are ubiquitous in transitional and turbulent shear flows. This is due to the lift-up effect: particle displaced in the cross-stream direction will initially retain their horizontal momentum and thus induce a perturbation in the streamwise velocity. Eventually viscosity will diffuse this momentum difference and damp the disturbance: however, in most cases, the disturbance is strong enough to generate strong localized shear layers and trigger new secondary instabilities.

\cite{ellingsen-Palm} showed that streamwise-independent perturbations in the cross-stream velocity will remain constant in the inviscid limit and induce a streamwise velocity disturbance that grows linearly in time. When the disturbance wave vector has an angle with the respect to the flow direction, 
the induced horizontal velocity disturbance resulting from the lift-up of a fluid
particle by the normal velocity is anyway such that the horizontal momentum in
the direction perpendicular to the wave number vector is conserved in the inviscid limit. The total amplification of the disturbance is however lower when the initial perturbation has a component in the flow direction. The optimal perturbation consists therefore of waves with wave vector orthogonal to the flow direction, i.e. streamwise-independent spanwise periodic waves.

The lift-up effect is identified has a key ingredient in subcritical transition in shear flows as well as one of the building blocks of wall-bounded turbulence, as shown in the first part of this review. 
Here we would like to conclude by noting an important point. 
Linear growth mechanisms are the only responsible for disturbance growth in shear flows, where nonlinear terms re-distribute energy and give zero net contribution when integrated over the whole control domain, see discussion about the Reynolds-Orr equation in \cite{Drazin81,Schmid2001}. 
\cite{dan:comment:nonormality:1996} discusses the role of linear terms in the transition and notes that, as a consequence of the Reynolds-Orr equation,  the growth rate of a
finite amplitude disturbance can, at each instant of its evolution, be found from an infinitesimal disturbance with identical
shape. Thus, the instantaneous growth rate of a finite
amplitude disturbance is given by mechanisms present in the 
linearized equations and the
growth of a finite amplitude disturbance can be regarded
as a sum of growth rates associated with the linear
mechanisms.
In shear flows, the main linear
mechanism for transient disturbance growth is the
lift-up effect that produces high and
low speed streaks in the streamwise velocity.

In the same spirit, the importance of linear processes in wall-bounded turbulent shear flows has been investigated through numerical experiments in \cite{jkim:jlim:2000}. It is shown that the linear coupling term in the Orr-Sommerfeld--Squire system, $-i\beta  U' \hat v $ in equation (\ref{eq:C4:OrrSquireMatrix}), responsible for the non-normality of the system, plays an important role also in fully turbulent flows. Near-wall turbulence indeed decays without the linear coupling term. It is also shown that near-wall turbulence structures are not formed in their proper scales without the nonlinear terms in the NavierÐStokes equations, thus indicating that the formation of the commonly observed near-wall turbulence structures are essentially nonlinear, but their maintenance relies on the linear process.
\cite{jimenez2013} discusses the importance of linear mechanisms in turbulence and in particular the role of Orr's inviscid mechanism in the transient amplification of disturbances
in shear flows in the context of bursting in the logarithmic layer of wall-bounded turbulence.
He shows how the nonlinear counterpart of the Orr mechanism \citep{orr:1907} is responsible for the regeneration of streamwise elongated structures, thus a step ahead of the lift-up effect inducing new streaks in the near-wall region. The disturbances
produced by the streak breakdown are amplified by an Orr-like transient process drawing
energy directly from the mean shear, rather than from the velocity gradients of
the streak.

In the second part of this work, we report recent and new results pertaining the lift-up effect in complex fluids, in particular non-Newtonian fluids, polymer suspensions and particle-laden channel flows. In these flows, we still identify the lift-up effect as the most dangerous mechanism for the disturbance growth. In the cases considered here, this is explained by the difference between the time scale over which streaks are formed and the time scale of the interactions between the flow and the suspended phase (polymer and particle relaxation times). Indeed, stabilization is observed for the growth of streamwise-dependent disturbances whose time evolution is faster than that of the elongated streaks.
In the future, it would be interesting to relate these modifications to the different dynamics of near-wall turbulence in particle-laden flows.
In particular, the case of finite-size particles may deserve further attention as the flow is in a chaotic state because of the continuous forcing by the particles.

To conclude, the lift-up effect is a relatively simple physical mechanisms leaving a permanent scar in wall-bounded shear flows.

\section*{Acknowledgements}

We are grateful to Dan Henningson, Carlo Cossu, Paolo Gualtieri, Stefania Cherubini and Tamer Zaki  for fruitful discussions.
Iman Lashgari, Mengqi Zhang, Joy Klinkenberg, Gaetano Sardina, Wim-Paul Breugem and Francesco Picano are acknowledged for helping in the analysis of viscoelastic and particle laden flows.
Computer time provided by SNIC (Swedish National Infrastructure
Centre) is acknowledged.

\bibliographystyle{elsart-harv}

\end{document}